\documentclass[final, twoside]{IEEEtran} 
\usepackage{float}
\usepackage{amssymb, amsmath, amsthm, bm}


\usepackage{graphicx, color}
\graphicspath{{figures-pdf/}}
\usepackage{pgfplots}[compat=newest]
\pgfplotsset{
tick label style={font=\small},
label style={font=\small},
legend style={font=\footnotesize, legend cell align=left, fill=none, draw=none},
major grid style={line width=0.2pt},
}

\usepackage{cite}
\usepackage{url}
\usepackage{diagbox}

\usepackage[aboveskip=1pt]{subcaption}
\usepackage[mathlines]{lineno}

\usepackage{algorithm}
\usepackage{algpseudocode}
\usepackage{multirow, bigstrut, threeparttable}
\usepackage{tabularx}
\usepackage{arydshln}
\usepackage{empheq}
\usepackage{datetime}
\usepackage{multirow}
\usepackage{booktabs}

\newlength\OneImW
\newlength\BigOneImW
\newlength\twofigwidth
\newlength\ThreeImW
\newlength\FourImW
\newlength\vfigskip
\usepackage{enumitem}
\setlist[itemize]{leftmargin=*}
\usepackage[bookmarks=false]{hyperref}
\hypersetup{
 linktocpage=true, pdfborderstyle={/S/S/W 1}, hyperindex=true, bookmarks=true, bookmarksopen=true, bookmarksnumbered=true,
}

\begin{document}

\title{Recovering the Block-wise Relationship in an Encryption-Then-Compression System}

\author{Chengqing Li,
Sheng Liu
\thanks{This work was supported by the National Natural Science Foundation of China (no.~92267102).}

\thanks{C. Li is with the School of Computer Science, Xiangtan University, Xiangtan 411105, Hunan, China (DrChengqingLi@gmail.com).}

\thanks{S. Liu is with the College of Computer Science and Electronic Engineering, Hunan University, Changsha 410082, Hunan, China.}
}

\markboth{IEEE Transactions}{Li~\MakeLowercase{et al.}}

\IEEEpubid{\begin{minipage}{\textwidth}\ \\[12pt] \centering
            1549-8328 \copyright 2019 IEEE. Personal use is permitted, but republication/redistribution requires IEEE permission.\\
            See http://www.ieee.org/publications standards/publications/rights/index.html for more information.
\end{minipage}}

\maketitle

\begin{abstract}
Joint encryption and compression is an ideal solution for protecting security and privacy of image data in a real scenario, e.g. storing them on an existing cloud-based service like Facebook. Recently, some block-wise encryption-then-compression (ETC) schemes compatible with JPEG were proposed to provide a reasonably high level of security without compromising compression ratio much. This paper investigates recovering the block-wise relationship in an ETC scheme exerting on single-color blocks of size $8\times 8$ in the scenarios of ciphertext-only attack, known-plaintext attack and chosen-plaintext attack. Then, the attacking targets are extended to the other conventional
ETC schemes exerting on multiple color channels and blocks of various sizes.
Especially, an elaborate jigsaw puzzle solver is designed to recover enough visual information from multiple cipher-images encrypted by the same secret key.
Moreover, the nice attacking performance was verified over two social media platforms, Facebook and Weibo.
\end{abstract}
\begin{IEEEkeywords}
Ciphertext-only attack, chosen-plaintext attack, image security, jigsaw puzzle solver, known-plaintext attack, permutation.
\end{IEEEkeywords}

\section{Introduction}

\IEEEPARstart{P}{rotecting} security and privacy of image data in cyberspace
is a constant uphill battle receiving everyone's concerns \cite{cqli:delay:IEEEM22}.
Meanwhile, due to massive capacity and high redundancy of image data,
compression is required to reduce image size and efficiently utilize network resources.
In recent years, with the assistance of some signal processing techniques, such as compressive sensing \cite{zhou:CS:TM21,cqli:CS:TMM23}, a variety of joint encryption and compression schemes were proposed to pursue a satisfying security level of image data meanwhile destroying its redundancy, the basis of compression, to a marginal extent
\cite{Wu:JCE:2005, Zhang:JCE:TMM2014, Zhang:JCE:TMM2016, He:JCE:TMM2018, Li:JCE:TMM2018, Shimizu:JCE:TIFS2021, Qin:JPEG:TMM2022, Li:JPEG:TITS2022, Singh:encrypt:2022}.

In certain scenarios, it is required to conduct image encryption prior to compression.
For example, a user wants to securely transmit images through an untrusted channel whose provider requires compressing the images to effectively exploit the constrained bandwidth and storage space.
To assure security or privacy, the user encrypts the images before transmission, and then the channel provider performs image processing in the encrypted domain.
It is well-known that image compression is tied to the high redundancy of image data and the strong correlation among neighboring pixels, whereas most encryption schemes eliminate the correlation and produce even completely random result.
Apparently, there is a contradiction between encryption and compression
when the former is conducted before the latter.
Therefore, the main challenge here is to minimize the size of image data without sacrificing security and privacy of users. In other words, the object is to seek for a balancing point in which both compression performance and security are acceptable for the given scenario
\cite{Tajik:thumbnail:NDSS19}.
To overcome the challenge, a number of image encryption schemes falling in the category of encryption-then-compression system (ETCS) were proposed in the past two decades
\cite{Johnson:OCED:TSP04, Schonberg:ETC:IP06, Schonberg:ETC:TIFS08, LiuW:ETC:TIP10, ZhangXP:ETC:TIFS11, ZhouJT:ETC:TIFS14, WangCT:ETC:SPIC15, Kumar:ETC:DSP17, Wang:ETC:TIFS18, Wang:ETC:CMS18, QinC:ETC:TCSVT19, Zhang:ETC:TMM2021}.

To endow ETCS with applicability for extensive scenarios, such as sharing images over an online social network (OSN) \cite{Zhoujt:share:TOMM18} and uploading images to a photo storage cloud \cite{Koh:cloud:Mobisys21},
the embedded compression scheme should be a widely-used standard.
To this end, a handful of block-wise perceptual encryption schemes compatible with JPEG compression standard or its variants were proposed in the past decade \cite{Kiya:ETCcoeff:ICASSP15, Kiya:ETCJPEG:PCS15, Kiya:ETCMJPEG:TFE15, Kiya:ETCJPEGXR:BMSB16, Kiya:ETCLossless:TIS17, Kiya:ETCRGB:SCIA17, Kiya:ETCRGB:TIS2018}.
In JPEG, the lossy compression is performed for each block separately.
Since almost all structural and statistical information within each block is preserved during the designed encryption process, the compression performance is only slightly reduced compared to that without any encryption process.
However, there exist two harsh drawbacks in ETCS proposed in \cite{Kiya:ETCJPEG:PCS15, Kiya:ETCMJPEG:TFE15, Kiya:ETCJPEGXR:BMSB16, Kiya:ETCLossless:TIS17}: restriction on block size and image distortion.
Since lossy JPEG compression on color images involves chrominance sub-sampling, the block size is limited to $16 \times 16$ to prevent severe degradation of compression performance, which may incur leaking local visual information and weakening security of the whole system.
In addition, if sub-sampling is used in compression, the decompression before decryption incurs severe distortion along  block edges.
Then, a ``grayscale-like" ETCS was proposed to strengthen security level of the whole system and eliminate the distortion by changing the encryption object from a three-color block to a single-color one, which naturally avoids the sub-sampling problem \cite{Kiya:ETC:TIFS19}.

\IEEEpubidadjcol 

In the aforementioned block-wise ETCS, the correlation among blocks in a plain-image is preserved in the encrypted result.
Due to the strong correlation existing in a natural image, the neighboring blocks of a given block can be found by searching for the blocks owning high correlation with it.
So, any attack on such category of systems utilizing the correlation
can be regarded as solving a jigsaw puzzle.
In \cite{Sholomon:puzzle:CVPR2013, Gallagher:puzzle:CVPR12, Paikin:puzzle:CVPR15, Son:puzzle:CVPR2016, Andalo:puzzle:TPAMI2016, Bridger:puzzle:CVPR2020},
the research on solving square jigsaw puzzle problem achieved great progress, and several powerful jigsaw puzzle solvers were proposed.
With the aid of the puzzle solvers, some ciphertext-only attacks were performed to evaluate the security performance of ETCS in \cite{Kiya:ETC:TIFS19, Kiya:safe:ICASSP17, Kiya:safe:ICME17, Kiya:safeDistorsion:IWSDA17, Kiya:safe:TIF18}.
But, the solver used in the attacks is not well-designed for puzzles introduced by ETCS and cannot handle the complicated encryption operations.
In \cite{Kiya:ETC:TIFS19}, according to some empirical attacking results, it was claimed that the grayscale-like ETCS owns enhanced security than the conventional ones, and can withstand ciphertext-only attack.
However, the statement is underpinned by an assumption of assigning a different secret key for each plain-image, which is impractical in a real scenario.
In \cite{Koh:cloud:Mobisys21}, a concrete implementation of ETCS is designed to protect users' photos stored on a cloud service, and in \cite{Kiya:DNN:IEEEM22}, ETCS is used to build a privacy-preserving deep learning framework, making objective evaluation of the real security performance of ETCS more important.

The security of grayscale-like ETCS was severely weakened in the pursuit of compression efficiency and format compliance. Its real security was much overestimated by the designers and the related followers.
ETCS may be only suitable for a certain scenario requiring security level much lower than the general expectation.
In this paper, we try to disclose the real security performance of ETCS against ciphertext-only attack,
known-plaintext attack and chosen-plaintext attack.
With the assistance of an elaborately designed jigsaw puzzle solver, we designed an efficient ciphertext-only attack method to reveal enough visual information from multiple cipher-images encrypted by the same secret key.
Furthermore, we tested the attacking performance in a real application scenario on sharing images over OSNs.
It is shown that even the available cipher-images underwent high-ratio lossy compression and other serious processing by OSNs,
the corresponding original images still can be recovered perfectly when the number of cipher-images is sufficient.
In plaintext attacks, we derived the equivalent secret key from the one-to-one correspondence between blocks in plain-images and cipher-images, which is determined by either using pixel-wise comparison or computing similarities among blocks.
For the former method, the correspondence is recovered by building a multi-branch tree, reducing the computational complexity remarkably.
To make the cryptanalysis more complete, the security evaluation of the conventional ETCS proposed in \cite{Kiya:ETCJPEG:PCS15, Kiya:ETCMJPEG:TFE15, Kiya:ETCJPEGXR:BMSB16, Kiya:ETCLossless:TIS17, Kiya:ETCRGB:SCIA17, Kiya:ETCRGB:TIS2018} is also presented with reference to that of ETCS proposed in \cite{Kiya:ETC:TIFS19}.

The rest of the paper is organized as follows. Section~\ref{sec:descrip} concisely introduces the procedure of ETCS.
Section~\ref{sec:COA} presents ciphertext-only attack on ETCS.
And known-plaintext attack and chosen-plaintext attack on ETCS are performed in Sec.~\ref{sec:PA}.
Then, Sec.~\ref{sec:conv} describes the cryptanalysis of a conventional ETCS.
The last section concludes the paper.

\section{Brief description of ETCS}
\label{sec:descrip}

The encryption object of ETCS proposed in \cite{Kiya:ETC:TIFS19} is an 8-bit RGB full-color image of size $W \times H$, which is denoted by $\mathbf{I}=\{I(i, j, k)\}_{i=0, j=0, k=0}^{W-1, H-1, 2}$.
The corresponding cipher-image is a grayscale image of size $3W\times H$ and can be represented as $\mathbf{I}'=\{I'(i, j)\}_{i=0, j=0}^{3W-1, H-1}$.
The basic parts of ETCS can be described as follows.
\begin{itemize}
\item \textit{The secret key}: three integers $k_1$, $k_2$, $k_3$ used for generating pseudorandom sequence.

\item \textit{Public parameter}: block size $W_{\rm B}$, $H_{\rm B}$.

\item \textit{Initialization}:
  \begin{itemize}
\item \textit{Step 1}: The plain-image first undergoes a color space conversion, where the RGB channels are transformed into YCbCr (Luminance, Chrominance blue, Chrominance red) triplets via
\begin{equation*}
      \left\{
      \begin{aligned}
        Y & = 0.299 \times R + 0.587 \times G + 0.114 \times B, \\
       Cb & = -0.1687 \times R - 0.3313 \times G + 0.5 \times B + 128, \\
       Cr & = 0.5 \times R - 0.4187 \times G - 0.0813 \times B + 128.
      \end{aligned}
      \right.
\end{equation*}
Denote the converted plain-image by $\mathbf{I}^*=\{I^*(i, j, k)\}_{i=0, j=0, k=0}^{W-1, H-1, 2}$.

\item \textit{Step 2}: Combine YCbCr triplets into one component $\mathbf{I}^{**}=\{I^{**}(i, j)\}_{i=0, j=0}^{3W-1, H-1}$, where $I^{**}(i+k*W, j)=I^*(i, j, k)$.
A plain-image ``Lenna" and its combination result are illustrated in Fig.~\ref{fig:plain256}.

\item \textit{Step 3}: Select $W_{\rm B}=H_{\rm B}=8$ as block size.
Divide $\mathbf{I}^{**}$ into $n$ non-overlapping blocks of size $W_{\rm B} \times H_{\rm B}$, where
\begin{equation*}
n=3\cdot \left\lfloor \frac{W}{W_{\rm B}}\right\rfloor \cdot \left\lfloor\frac{H}{H_{\rm B}}\right\rfloor.
\end{equation*}
The blocks are scanned in the raster order and then represented as $\{B(i)\}_{i=0}^{n-1}$.
Without loss of generality, assume that $W$ and $H$ can be exactly divided by $W_{\rm B}$ and $H_{\rm B}$, respectively.
Then generate three pseudorandom integer sequences $\{s(i)\}_{i=0}^{n-1}$, $\{(r(i), f(i))\}_{i=0}^{n-1}$, and $\{t(i)\}_{i=0}^{n-1}$
using the subkeys $k_1$, $k_2$, and $k_3$, respectively, where $0 \leq s(i) \leq n-1$, $0 \leq r(i), f(i) \leq 3$, and $t(i)\in \{0, 1\}$ \footnote{The concrete PRNG is not mentioned in \cite{Kiya:ETC:TIFS19}. Fortunately, this is not related with the insecurity problems studied in this paper.}.
\end{itemize}

\begin{figure}[!htb]
 \centering
 \begin{minipage}{0.33\BigOneImW}
   \centering
   \includegraphics[width=0.33\BigOneImW]{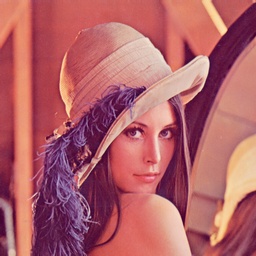}
   a)
 \end{minipage} \hspace{\vfigskip}
 \begin{minipage}{\BigOneImW}
   \centering
   \includegraphics[width=\BigOneImW]{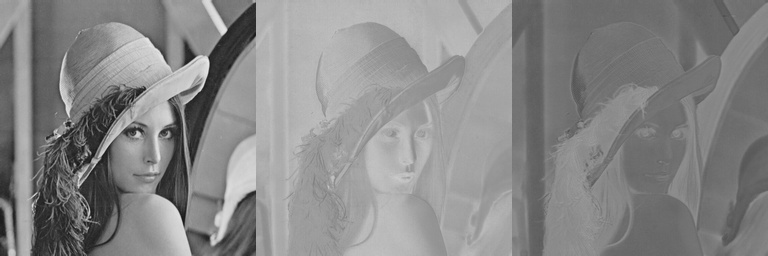}
   b)
 \end{minipage}
\caption{A plain-image and its combination result: a) a color image; b) the flatten grayscale version of a).}
\label{fig:plain256}
\end{figure}

\item \textit{The encryption procedure}:

\begin{itemize}
\item \textit{Step 1}: Permute blocks $\{B(i)\}_{i=0}^{n-1}$ randomly using the sequence $\{s(i)\}_{i=0}^{n-1}$, where $s(i)$ is used to swap the locations of blocks $B(i)$ and $B(s(i))$.

\item \textit{Step 2}: Rotate and invert blocks using $\{(r(i), f(i))\}_{i=0}^{n-1}$.
For each block $B(i)$, the four possible values of $r(i)$ indicates rotating it $0^\circ$, $90^\circ$, $180^\circ$, and $270^\circ$, respectively.
Similarly, $f(i)$ denotes four possible inverting directions: horizontal, vertical, both, and neither.

\item \textit{Step 3}: Perform negative-positive transformation (NPT) on every block using $\{t(i)\}_{i=0}^{n-1}$.
For each $L$-bit pixel $p$ in block $B(i)$, calculate
\begin{equation*}
p'=
\begin{cases}
p                  & \text{if } t(i)=0;\\
p \oplus( 2^L-1)   & \text{if } t(i)=1,
\end{cases}
\end{equation*}
where $\oplus$ denotes the bitwise XOR operation.
\end{itemize}

\item \textit{Compression}: Apply JPEG compression to the cipher-image.
\end{itemize}

\section{Ciphertext-only attack on ETCS}
\label{sec:COA}

Due to the special properties of the basic operations of ETCS,
the correlation among divided blocks remains almost intact after the block-wise encryption.
Even in the scenario of ciphertext-only attack, one can regard reconstructing the plain-image from the cipher-image as a jigsaw puzzle.
In \cite{Kiya:ETC:TIFS19, Kiya:safe:ICASSP17, Kiya:safe:ICME17, Kiya:safeDistorsion:IWSDA17, Kiya:safe:TIF18}, Kiya {\em et al.} performed ciphertext-only attacks on ETCS with the aid of a jigsaw puzzle solver.
In \cite{Kiya:ETC:TIFS19}, it is claimed that ETCS ``becomes robust against known-plaintext attack through the assigning of a different key to each image for the encryption". This means that ETCS is similar to a one-time password (also known as a one-time authorization code), which is impractical as complex key management is required to avoid any repeating usage of secret keys.
To keep such management successful, many loads on human resources and computation would be incurred and
satisfying balancing point between security and usability cannot be achieved.
Note that most symmetric encryption standards, e.g. DES, IDEA, permit repeated usage of a secret key for a relatively long time. In \cite{Kiya:ETC:TIFS19}, no concrete implementation detail on key management is mentioned.
So, we cannot analyze its security from the perspective of system security and/or human-centered security.
In all, the assumption of ETCS on using every secret key only one time is practically unreasonable.

Solving a jigsaw puzzle involves two basic problems:
1) judging whether two transformed blocks are neighboring to each other before encryption;
2) reassemble the scrambled blocks in the order approaching the original one as much as possible.
The kernels of the two problems lie in pairwise compatibility metric and jigsaw puzzle solver, respectively.
For convenience, we term the special jigsaw puzzle introduced by ETCS and that involves block permutation and rotation only as Type 1 puzzle and Type 0 puzzle.
In this section, we first discuss the pairwise compatibility metric in the context of cryptanalysis, which has distinctive characteristics compared to that used in an ordinary jigsaw puzzle.
Then, we elaborate a sophisticated jigsaw puzzle solver tailored to Type 1 puzzle.
Finally, a number of experiments are performed over various conditions, including a real-life scenario of sharing images via OSNs, to demonstrate satisfying performance of the proposed attacking method.

\subsection{Pairwise compatibility}
\label{sec:COA:pc}

A pairwise compatibility metric is used to estimate the possibility that two blocks abut to each other in a given setting, for example, one square block is placed in one direction (left, right, top or bottom) of another square block.
The compatibility is normally calculated using the boundary pixel values.
Simple metric SSD (Sum of Square Distance) sums up the square differences of pixels along the adjacent boundary.
To grasp the influence of gradient change of pixel values for compatibility, Gallagher introduced metric MGC (Mahalanobis Gradient Compatibility) \cite{Gallagher:puzzle:CVPR12}.
Son {\em et al.} integrated the changes of pixels along the boundary within the block itself, which can be considered as an enhanced version of MGC (EMGC) \cite{Son:puzzle:CVPR2016}.
However, the above metrics all become unreliable and inaccurate when the size of blocks gets relatively small, in which the existing puzzle solvers cannot produce acceptable result.
This is also the basis of the statement on the high security of ETCS given in \cite{Kiya:ETC:TIFS19, Kiya:safe:ICASSP17, Kiya:safe:ICME17, Kiya:safeDistorsion:IWSDA17, Kiya:safe:TIF18}.
The above compatibility metrics can work better when multiple cipher-images encrypted by the same secret key are available. Note that such assumption is reasonable as reusing secret keys is acceptable and necessary as mentioned before.
Combining enough permuted blocks, the existing compatibility metrics can produce accurate result even for small-size blocks.

\begin{figure}[!htb]
  \centering
  \begin{minipage}{\twofigwidth}
    \centering
    \includegraphics[width=\twofigwidth]{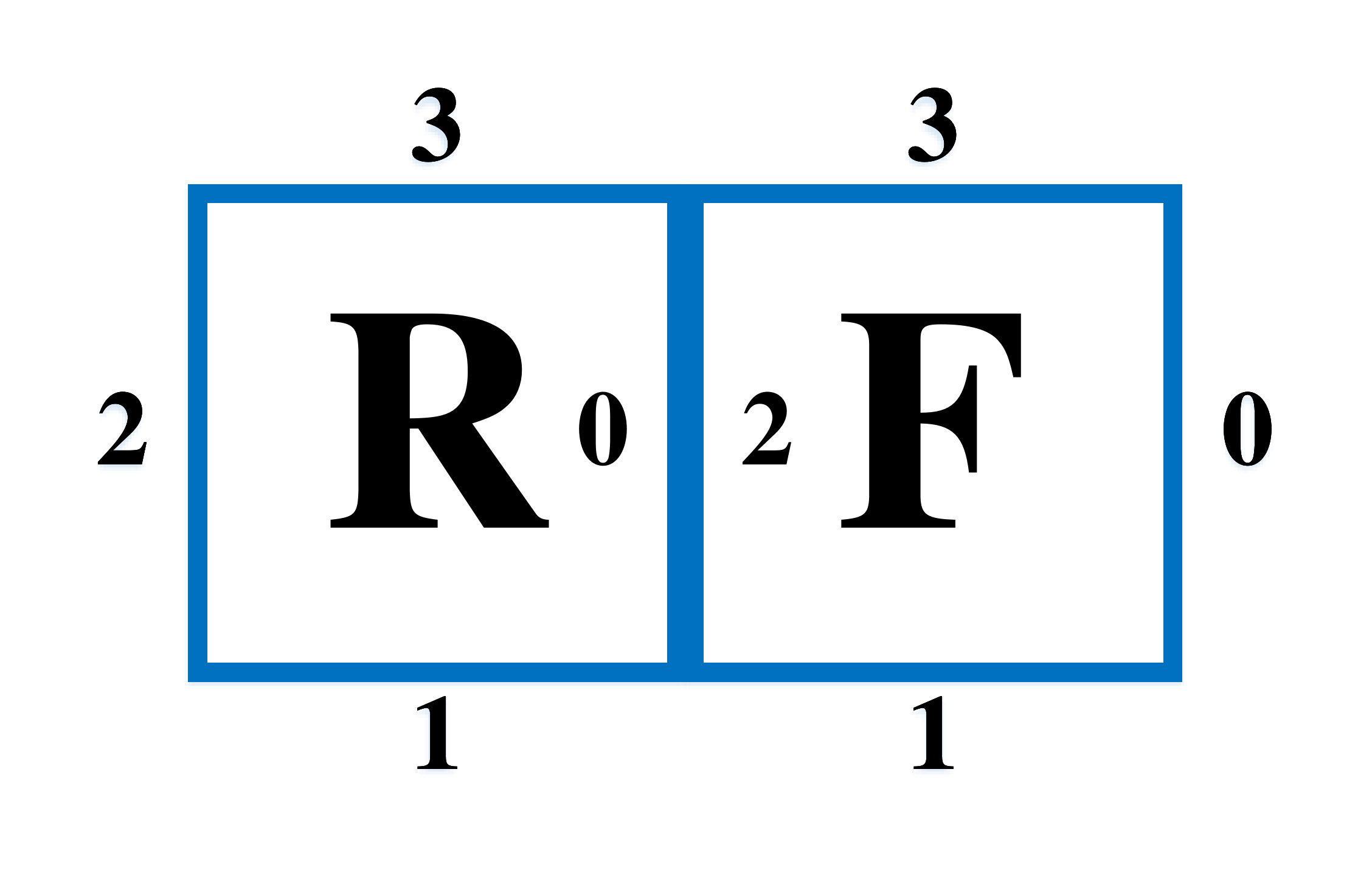}
    a)
  \end{minipage}
  \begin{minipage}{\twofigwidth}
    \centering
    \includegraphics[width=\twofigwidth]{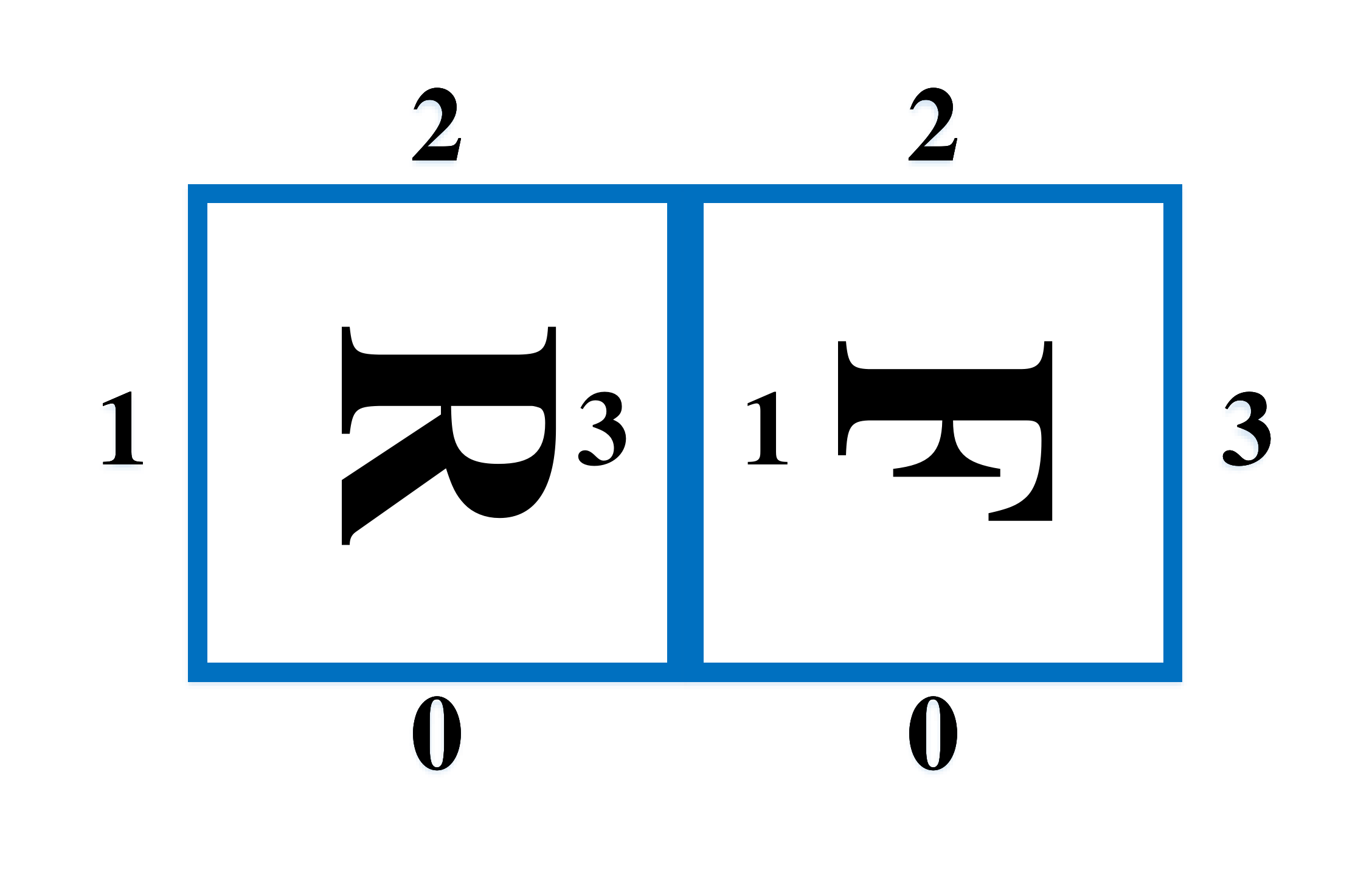}
    b)
  \end{minipage}
  \begin{minipage}{\twofigwidth}
    \centering
    \includegraphics[width=\twofigwidth]{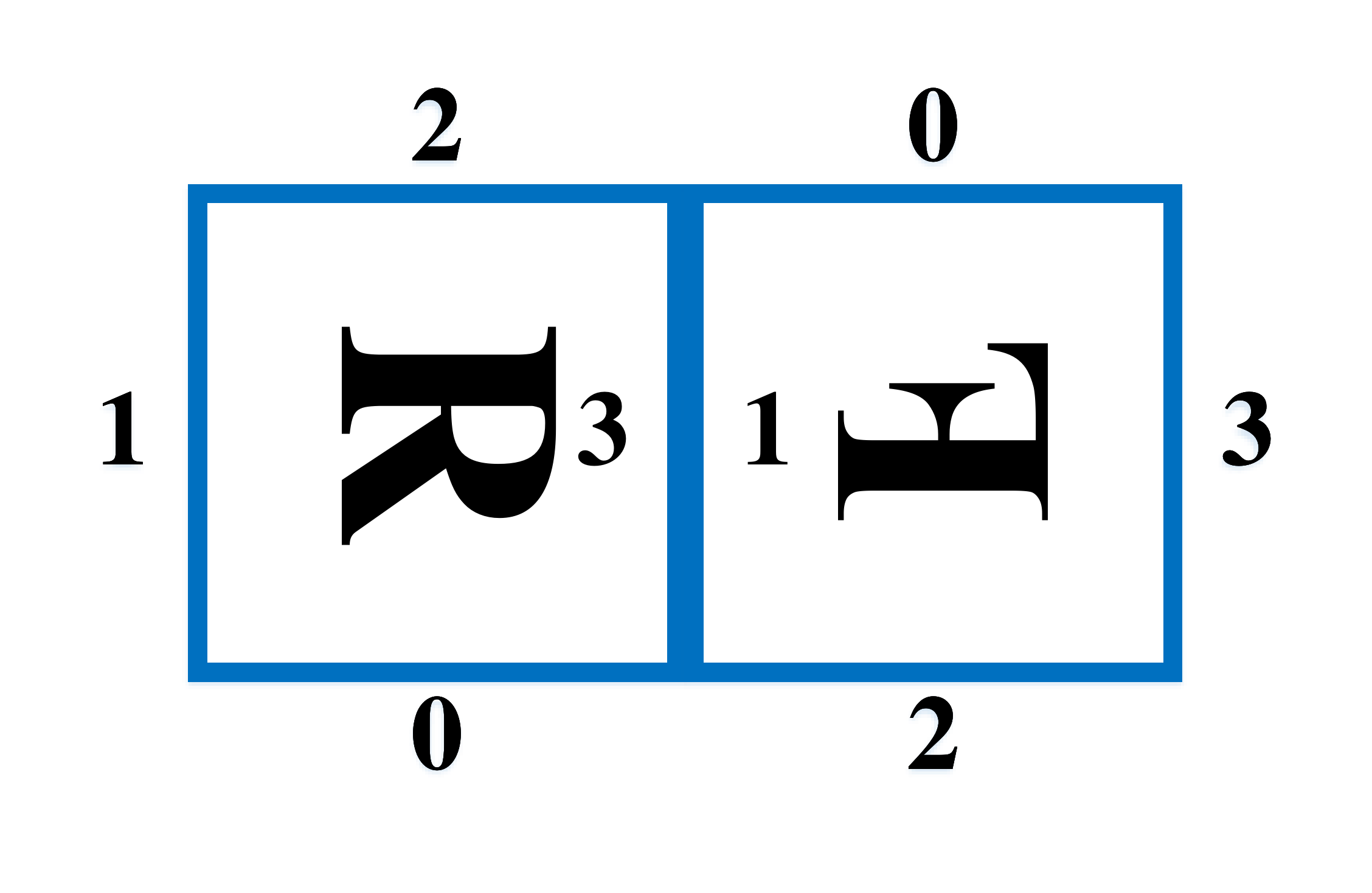}
    c)
  \end{minipage}
  \begin{minipage}{\twofigwidth}
    \centering
    \includegraphics[width=\twofigwidth]{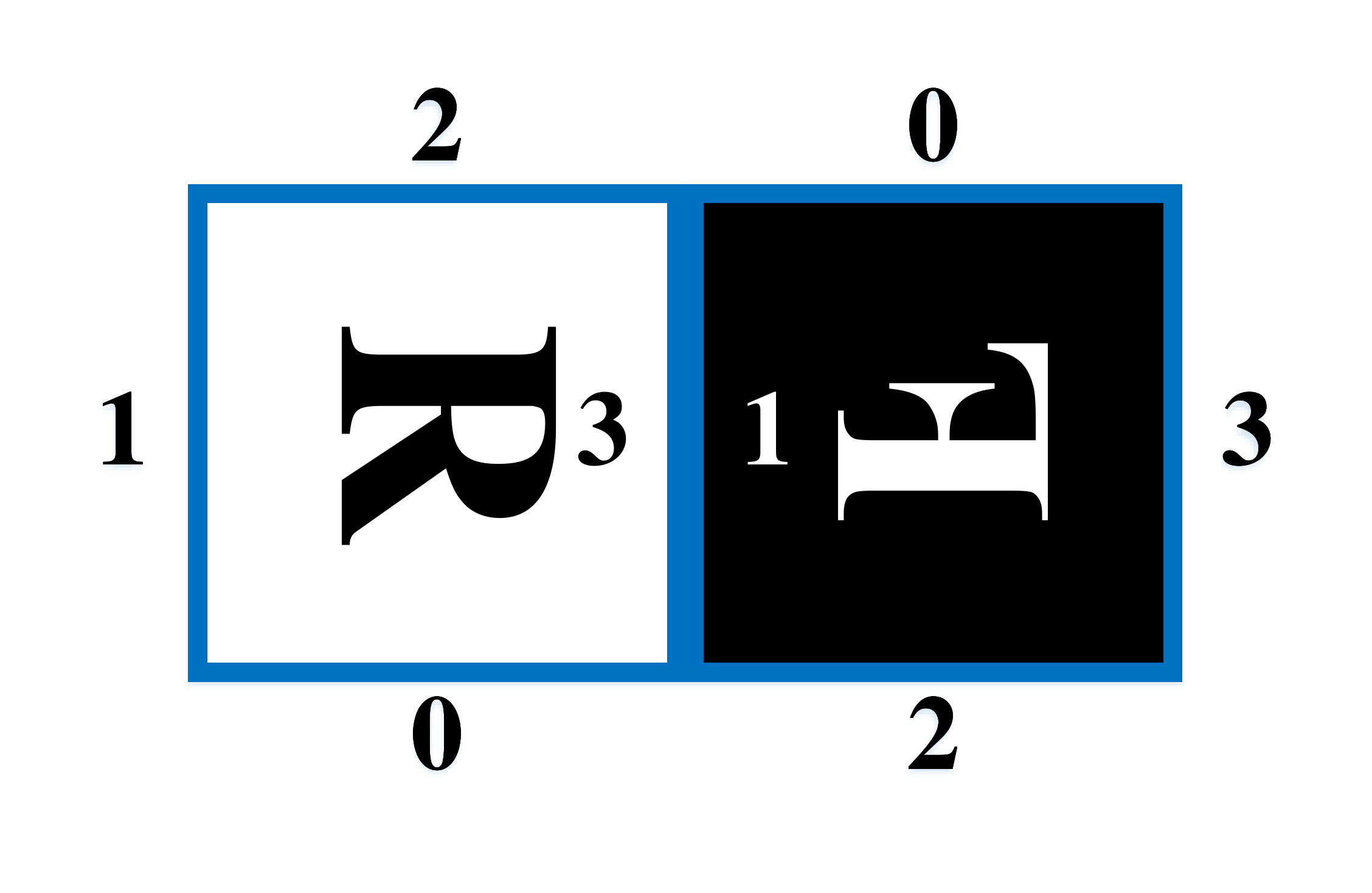}
    d)
  \end{minipage}
\caption{Four different configurations on connecting blocks $``R"$ and $``F"$ together:
 a) $(R, F, 0, 2, 0, 0)$;
 b) $(R, F, 3, 1, 0, 0)$;
 c) $(R, F, 3, 1, 1, 0)$;
 d) $(R, F, 3, 1, 1, 1)$.}
\label{fig:block}
\end{figure}

To calculate the compatibility metrics between the mapping results of two blocks in the encrypted domain, we have to encode their relative position first.
Given block $u$, let $i_u$ and $t_u$ denote whether to apply inversion and NPT to block $u$, respectively, where $i_u, t_u\in \{0, 1\}$.
Similarly, let $i_v$ and $u_v$ signal the corresponding operations for block $v$.
Then, the 64 possible configurations between blocks $u$ and $v$ can be unambiguously denoted by $e=(u, v, s_u, s_v, i, t)$, where $s_u, s_v \in[0,3]$ denote the sides of $u$ and $v$, respectively, $i=i_u\oplus i_v$, and $t=t_u\oplus t_v$.
The relationship represented by $e$ can also be regarded as a match between $u$ and $v$.
Some examples of different matches are illustrated in Fig.~\ref{fig:block}, where the sides of blocks are numbered for better visualization.

\graphicspath{{figures-pdf/block/}}

The compatibility among all blocks in a cipher-image can be stored in an array of size $n\times 4\times n\times 16$.
After calculating compatibility score for each cipher-image, we average their counterparts. Then, for each side of every block, we divide its $n\times 16$ compatibility scores by the second smallest one \cite{Gallagher:puzzle:CVPR12}.
In a Type 0 puzzle, there are 16 configurations connecting two blocks together.
Given a function of calculating the compatibility between two blocks placed together horizontally, as shown in Fig.~\ref{fig:block}a), $4\times 4=16$ distinct compatibility scores can be calculated by separately rotating the two blocks four possible degrees: $0^\circ$, $90^\circ$, $180^\circ$, or $270^\circ$.
For the configurations, applying vertically inversion and/or NPT to the right blocks multiplies the number of compatibility four times in Type 1 puzzle, as shown in Fig.~\ref{fig:block}c) and d). Note that manipulating the left block does not generate a new combination.

\begin{figure}[!htb]
\centering
\subfloat[$128\times 128$]{\pgfplotstableread{
 n   ssd      mgc      emgc
 1 0.152431 0.170486 0.149653
 2 0.338889 0.332639 0.294792
 3 0.482292 0.475694 0.414931
 4 0.515625 0.509028 0.446528
 5 0.651389 0.660764 0.606944
 6 0.707292 0.720486 0.664236
 7 0.743750 0.760764 0.713542
 8 0.794097 0.795833 0.757986
 9 0.835069 0.843056 0.811458
10 0.860069 0.863889 0.841319
11 0.878819 0.886806 0.864583
12 0.901389 0.906944 0.885069
13 0.908681 0.915972 0.896875
14 0.928819 0.934028 0.916319
15 0.935764 0.938889 0.924306
16 0.941667 0.944792 0.933333
17 0.946181 0.948264 0.936458
18 0.949306 0.952083 0.942708
19 0.957986 0.960069 0.952083
20 0.963889 0.966319 0.961806
}\loadedtable
\begin{tikzpicture}[scale=0.5]
\begin{axis}[
xlabel={Number of cipher-images ($M$)},
ylabel={Accuracy},
xmin=1, xmax=20,
xtick={1, 4, 8, 12, 16, 20},
ymin=0, ymax=1,
ytick={0, 0.1, ..., 1.01},
tick label style={font=\large},
label style={font=\Large},
legend style={font=\large},
legend pos=south east,
xmajorgrids, ymajorgrids,
]
\addplot [red, mark=*] table[x=n, y=ssd] {\loadedtable};
\addlegendentry{SSD}

\addplot [green, mark=square] table[x=n, y=mgc]{\loadedtable};
\addlegendentry{MGC}

\addplot [blue, mark=x] table[x=n, y=emgc]{\loadedtable};
\addlegendentry{EMGC}

\end{axis}
\end{tikzpicture}}
\thinspace
\subfloat[$256\times 256$]{\pgfplotstableread{
 n   ssd      mgc      emgc
 1 0.069136 0.078545 0.074933
 2 0.238575 0.261425 0.234795
 3 0.395329 0.434056 0.397681
 4 0.433300 0.476058 0.433972
 5 0.587450 0.644993 0.593918
 6 0.646337 0.700941 0.645161
 7 0.685316 0.737315 0.687584
 8 0.742860 0.781922 0.740087
 9 0.801915 0.833585 0.804519
10 0.826529 0.853411 0.825689
11 0.850218 0.875000 0.850134
12 0.865423 0.886929 0.863575
13 0.873068 0.892221 0.870884
14 0.886509 0.906082 0.887685
15 0.898101 0.917255 0.902302
16 0.909358 0.926915 0.914231
17 0.913222 0.930024 0.918263
18 0.917507 0.933384 0.921875
19 0.923807 0.939600 0.931368
20 0.928091 0.941196 0.936744
}\loadedtable
\begin{tikzpicture}[scale=0.5]
\begin{axis}[
xlabel={Number of cipher-images ($M$)},
ylabel={Accuracy},
xmin=1, xmax=20,
xtick={1, 4, 8, 12, 16, 20},
ymin=0, ymax=1,
ytick={0, 0.1, ..., 1.01},
tick label style={font=\large},
label style={font=\Large},
legend style={font=\large},
legend pos=south east,
xmajorgrids, ymajorgrids,
]
\addplot [red, mark=*] table[x=n, y=ssd] {\loadedtable};
\addlegendentry{SSD}

\addplot [green, mark=square] table[x=n, y=mgc]{\loadedtable};
\addlegendentry{MGC}

\addplot [blue, mark=x] table[x=n, y=emgc]{\loadedtable};
\addlegendentry{EMGC}

\end{axis}
\end{tikzpicture}}
\quad
\subfloat[$512\times 512$]{\pgfplotstableread{
 n   ssd      mgc      emgc
 1 0.026972 0.032903 0.030878
 2 0.118448 0.120784 0.101087
 3 0.365617 0.400876 0.347946
 4 0.591663 0.628431 0.562066
 5 0.697607 0.720155 0.671627
 6 0.761347 0.782759 0.734086
 7 0.780506 0.792679 0.735346
 8 0.746052 0.741774 0.657676
 9 0.778956 0.775318 0.700438
10 0.809979 0.809317 0.740493
11 0.830171 0.830254 0.768498
12 0.858817 0.858920 0.805514
13 0.863757 0.863839 0.815187
14 0.873905 0.873719 0.826802
15 0.877749 0.877232 0.832238
16 0.894614 0.893849 0.852741
17 0.900752 0.899843 0.861669
18 0.919416 0.926298 0.898830
19 0.925285 0.931403 0.905651
20 0.935144 0.940414 0.917245
}\loadedtable
\begin{tikzpicture}[scale=0.5]
\begin{axis}[
xlabel={Number of cipher-images ($M$)},
ylabel={Accuracy},
xmin=1, xmax=20,
xtick={1, 4, 8, 12, 16, 20},
ymin=0, ymax=1,
ytick={0, 0.1, ..., 1.01},
tick label style={font=\large},
label style={font=\Large},
legend style={font=\large},
legend pos=south east,
xmajorgrids, ymajorgrids,
]
\addplot [red, mark=*] table[x=n, y=ssd] {\loadedtable};
\addlegendentry{SSD}

\addplot [green, mark=square] table[x=n, y=mgc]{\loadedtable};
\addlegendentry{MGC}

\addplot [blue, mark=x] table[x=n, y=emgc]{\loadedtable};
\addlegendentry{EMGC}

\end{axis}
\end{tikzpicture}}
\thinspace
\subfloat[$1024\times 1024$]{\pgfplotstableread{
 n   ssd      mgc      emgc
 1 0.008822 0.011903 0.012026
 2 0.133699 0.169071 0.150503
 3 0.462470 0.529794 0.481058
 4 0.652251 0.704027 0.665103
 5 0.744289 0.786669 0.751456
 6 0.762190 0.796147 0.755762
 7 0.650739 0.658747 0.582216
 8 0.695333 0.703909 0.636057
 9 0.742726 0.751348 0.685301
10 0.775232 0.784157 0.721164
11 0.815965 0.824332 0.769070
12 0.823855 0.831847 0.780799
13 0.836312 0.844509 0.795260
14 0.841141 0.849538 0.801889
15 0.859288 0.867121 0.822619
16 0.867408 0.875077 0.832323
17 0.895269 0.909326 0.875651
18 0.900975 0.915108 0.882402
19 0.914765 0.927099 0.897879
20 0.918779 0.932651 0.906978
}\loadedtable
\begin{tikzpicture}[scale=0.5]
\begin{axis}[
xlabel={Number of cipher-images ($M$)},
ylabel={Accuracy},
xmin=1, xmax=20,
xtick={1, 4, 8, 12, 16, 20},
ymin=0, ymax=1,
ytick={0, 0.1, ..., 1.01},
tick label style={font=\large},
label style={font=\Large},
legend style={font=\large},
legend pos=south east,
xmajorgrids, ymajorgrids,
]
\addplot [red, mark=*] table[x=n, y=ssd] {\loadedtable};
\addlegendentry{SSD}

\addplot [green, mark=square] table[x=n, y=mgc]{\loadedtable};
\addlegendentry{MGC}

\addplot [blue, mark=x] table[x=n, y=emgc]{\loadedtable};
\addlegendentry{EMGC}

\end{axis}
\end{tikzpicture}}
\caption{Accuracy of three compatibility metrics when different number of cipher-images of various sizes are available.}
\label{fig:metric}
\end{figure}
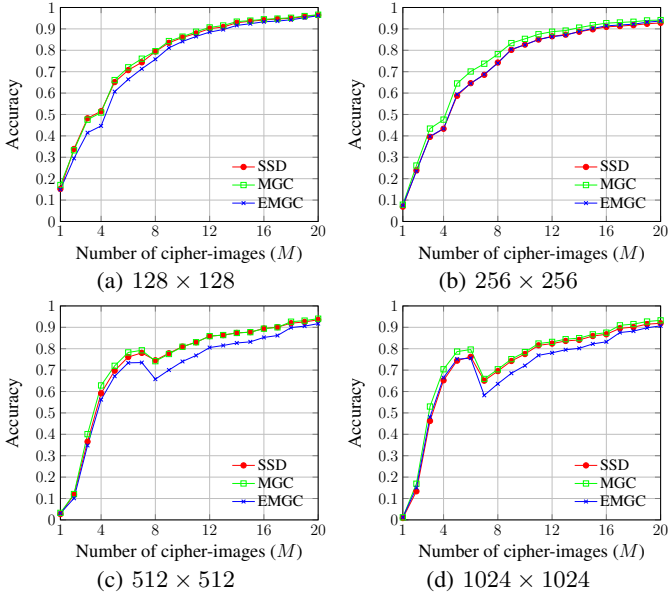

When solving puzzles, we care about whether the correct matches can be found from all potential ones by these metrics.
This can be evaluated by the ratio of the correct matches that own the highest compatibility score compared to the false others.
As shown in Fig.~\ref{fig:metric}, with the increase of the number of cipher-images, three representative metrics, SSD, MGC, and EMGC, can all confirm adequate correct matches.
Even for cipher-images of larger sizes (containing more blocks), they also demonstrate stable performance, e.g.
the ratio is larger than 0.9 when the number of available
cipher-images is larger than 20.
Thus, we choose plain-images of size $256\times 256$ in the subsequent experiments if not otherwise specified.
Moreover, we select MGC to solve Type 1 puzzles considering computation complexity and effectiveness.

\begin{figure}[!htb]
\centering
\pgfplotstableread{
 n pQ       eQ       pQQ      eQQ      pQQQ     eQQQ
 1 0.121724 0.078545 0.064852 0.035786 0.043263 0.022597
 2 0.332325 0.261425 0.195649 0.139533 0.139029 0.096522
 3 0.500168 0.434056 0.310988 0.242944 0.219170 0.163558
 4 0.537718 0.476058 0.350134 0.280830 0.255460 0.202201
 5 0.697581 0.644993 0.512517 0.444724 0.398017 0.330981
 6 0.747228 0.700941 0.577033 0.509913 0.462534 0.392389
 7 0.781082 0.737315 0.622648 0.554267 0.510585 0.436072
 8 0.820060 0.781922 0.678931 0.613155 0.568548 0.497144
 9 0.865171 0.833585 0.768733 0.713122 0.672883 0.605007
10 0.881972 0.853411 0.798891 0.747648 0.708333 0.643313
11 0.898185 0.875000 0.826865 0.776546 0.740423 0.677167
12 0.909526 0.886929 0.849126 0.803847 0.768901 0.708165
13 0.913726 0.892221 0.860971 0.817372 0.782930 0.724546
14 0.925823 0.906082 0.879452 0.838710 0.808720 0.751680
15 0.934476 0.917255 0.892893 0.853831 0.824513 0.770413
16 0.942876 0.926915 0.905410 0.869288 0.840894 0.793599
17 0.945481 0.930024 0.908686 0.875084 0.848118 0.802839
18 0.948421 0.933384 0.912970 0.880124 0.853075 0.808636
19 0.953713 0.939600 0.923639 0.894825 0.868952 0.827537
20 0.955393 0.941196 0.927671 0.901294 0.878024 0.836526
21 0.956485 0.942204 0.929267 0.902722 0.878948 0.838458
22 0.962030 0.949009 0.940524 0.917591 0.901882 0.865927
23 0.963962 0.952789 0.944724 0.925991 0.909274 0.878276
24 0.968246 0.957493 0.952117 0.935568 0.920867 0.890457
25 0.969590 0.959677 0.954553 0.939096 0.926579 0.897009
26 0.969842 0.959929 0.955645 0.940188 0.927671 0.898942
27 0.973034 0.964382 0.960601 0.946321 0.934056 0.908602
28 0.971942 0.964970 0.960601 0.948505 0.936240 0.914819
29 0.972614 0.965726 0.961358 0.949093 0.937668 0.916751
30 0.974210 0.966230 0.964130 0.952369 0.941532 0.921539
31 0.975050 0.967154 0.966482 0.954553 0.946069 0.925907
32 0.976815 0.969254 0.968414 0.956737 0.947833 0.928259
}\loadedtable
\begin{tikzpicture}[scale=0.85]
\begin{axis}[
xlabel={Number of cipher-images ($M$)},
ylabel={Accuracy},
xmin=1, xmax=32,
xtick={1, 4, 8, 12, 16, 20, 24, 28, 32},
ymin=0, ymax=1,
ytick={0, 0.1, ..., 1.01},
legend pos=south east,
xmajorgrids, ymajorgrids,
]

\addplot [red, mark=square] table[x=n, y=pQ] {\loadedtable};
\addlegendentry{Type 1 ($Q=95$)}
\addplot [red, mark=triangle] table[x=n, y=pQQ] {\loadedtable};
\addlegendentry{Type 1 ($Q=71$)}
\addplot [red, mark=o] table[x=n, y=pQQQ] {\loadedtable};
\addlegendentry{Type 1 ($Q=47$)}

\addplot [densely dotted, cyan, mark=x, mark options=solid] table[x=n, y=eQ] {\loadedtable};
\addlegendentry{Type 0 ($Q=95$)}
\addplot [densely dotted, cyan, mark=+, mark options=solid] table[x=n, y=eQQ] {\loadedtable};
\addlegendentry{Type 0 ($Q=71$)}
\addplot [densely dotted, cyan, mark=10-pointed star, mark options=solid] table[x=n, y=eQQQ] {\loadedtable};
\addlegendentry{Type 0 ($Q=47$)}

\end{axis}
\end{tikzpicture}
\caption{Security comparison between two types of jigsaw puzzle.}
\label{fig:mgc}
\end{figure}
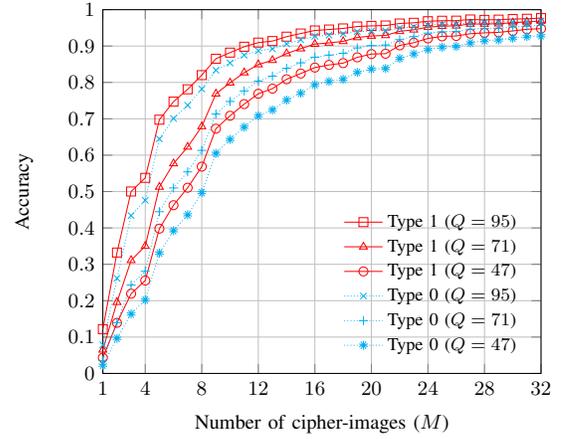

Similar to ETCS, Type 0 puzzle can also be used to construct an image encryption scheme, in which blocks are permuted and rotated randomly.
Hence, we can examine the security improvement of ETCS by comparing the accuracies of MGC regarding Type 1 and 0 puzzles.
As depicted in Fig.~\ref{fig:mgc}, their accuracies are very close under the same JPEG quality factor $Q$.
Therefore, there is only a slight improvement in terms of security via introducing inversion and NPT operations in ETCS.
In addition, the accuracy gap between different compression levels becomes narrow when more cipher-images are adopted.

\subsection{Jigsaw puzzle solver}
\label{ssec:jps}

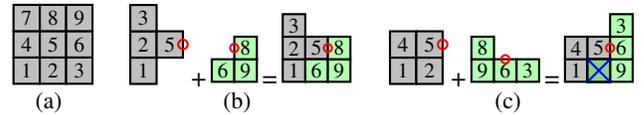
\begin{figure}[!htb]
\centering
\subfloat[]{\begin{tikzpicture}[thick=ultra thick, scale=0.35, font=\footnotesize]
\foreach \y in {0, 1, 2}
\foreach \x in {0, 1, 2} {
  \pgfmathsetmacro\zigzagid{int(\y*3+\x+1)}
  \filldraw[fill=lightgray] (\x,\y) rectangle (\x+1,\y+1);
  \node at (\x+0.5, \y+0.5) {\zigzagid};
}
\end{tikzpicture}}
\quad
\subfloat[]{\begin{tikzpicture}[thick=ultra thick, scale=0.35, font=\footnotesize]
\foreach \y in {1, 2, 3}
{
  \filldraw[fill=lightgray] (0,\y-1) rectangle (1,\y);
  \node at (0.5, \y-0.5) {\y};
}
\filldraw[fill=lightgray] (1,1) rectangle (2,2);
\node at (1.5, 1.5) {5};
\draw[red] (2, 1.5) circle [radius=0.18];
\end{tikzpicture}
\hspace{-0.25em}+\hspace{-0.25em}
\begin{tikzpicture}[thick=ultra thick, scale=0.3, font=\footnotesize]
\filldraw[fill=green!30] (0,0) rectangle (1,1);
\node at (0.5, 0.5) {6};
\filldraw[fill=green!30] (1,0) rectangle (2,1);
\node at (1.5, 0.5) {9};
\filldraw[fill=green!30] (1,1) rectangle (2,2);
\node at (1.5, 1.5) {8};
\draw[red] (1, 1.5) circle [radius=0.18];
\end{tikzpicture}
\hspace{-0.25em}=\hspace{-0.25em}
\begin{tikzpicture}[thick=ultra thick, scale=0.3, font=\footnotesize]
\foreach \y in {1, 2, 3} {
  \filldraw[fill=lightgray] (0,\y-1) rectangle (1,\y);
  \node at (0.5, \y-0.5) {\y};
}
\filldraw[fill=lightgray] (1,1) rectangle (2,2);
\node at (1.5, 1.5) {5};

\filldraw[fill=green!30] (1,0) rectangle (2,1);
\node at (1.5, 0.5) {6};
\filldraw[fill=green!30] (2,0) rectangle (3,1);
\node at (2.5, 0.5) {9};
\filldraw[fill=green!30] (2,1) rectangle (3,2);
\node at (2.5, 1.5) {8};

\draw[red] (2, 1.5) circle [radius=0.18];

\end{tikzpicture}}
\quad
\subfloat[]{\begin{tikzpicture}[thick=ultra thick, scale=0.35, font=\footnotesize]
\foreach \x in {1, 2} {
  \filldraw[fill=lightgray] (\x-1,0) rectangle (\x,1);
  \node at (\x-0.5, 0.5) {\x};
}
\foreach \x in {4, 5} {
  \filldraw[fill=lightgray] (\x-4,1) rectangle (\x-3,2);
  \node at (\x-3.5, 1.5) {\x};
}
\draw[red] (2, 1.5) circle [radius=0.18];
\end{tikzpicture}
\hspace{-0.25em}+\hspace{-0.25em}
\begin{tikzpicture}[thick=ultra thick, scale=0.3, font=\footnotesize]
\filldraw[fill=green!30] (0,0) rectangle (1,1);
\node at (0.5, 0.5) {9};
\filldraw[fill=green!30] (1,0) rectangle (2,1);
\node at (1.5, 0.5) {6};
\filldraw[fill=green!30] (2,0) rectangle (3,1);
\node at (2.5, 0.5) {3};
\filldraw[fill=green!30] (0,1) rectangle (1,2);
\node at (0.5, 1.5) {8};
\draw[red] (1.5, 1) circle [radius=0.18];
\end{tikzpicture}
\hspace{-0.25em}=\hspace{-0.25em}
\begin{tikzpicture}[thick=ultra thick, scale=0.3, font=\footnotesize]
\foreach \x in {1, 2} {
  \filldraw[fill=lightgray] (\x-1,0) rectangle (\x,1);
  \node at (\x-0.5, 0.5) {\x};
}
\foreach \x in {4, 5} {
  \filldraw[fill=lightgray] (\x-4,1) rectangle (\x-3,2);
  \node at (\x-3.5, 1.5) {\x};
}

\filldraw[fill=green!30] (1,0) rectangle (2,1);
\node at (1.5, 0.5) {8};
\filldraw[fill=green!30] (2,0) rectangle (3,1);
\node at (2.5, 0.5) {9};
\filldraw[fill=green!30] (2,1) rectangle (3,2);
\node at (2.5, 1.5) {6};
\filldraw[fill=green!30] (2,2) rectangle (3,3);
\node at (2.5, 2.5) {3};
\draw[red] (2, 1.5) circle [radius=0.18];

\filldraw[fill=green!30!lightgray] (1,0) rectangle (2,1);
\draw[blue] (1,0) -- (2,1);
\draw[blue] (1,1) -- (2,0);
\end{tikzpicture}}
\caption{Serious mistakes incurred when solving Type 1 puzzle by Gallagher's puzzle solver: a) The original $3\times3$ jigsaw puzzle; b) False merging result; c) Merging fails as the collision between blocks marked with ``X", where the sides to be joined together are marked with circles.}
\label{fig:error}
\end{figure}

In the past decade, some powerful jigsaw puzzle solver were proposed, which can deal well with up to 22,755 blocks \cite{Sholomon:puzzle:CVPR2013}, mixed blocks \cite{Gallagher:puzzle:CVPR12}, missing blocks \cite{Paikin:puzzle:CVPR15}, eroded boundaries \cite{Bridger:puzzle:CVPR2020}.
However, they were originally designed for Type 0 puzzle that only involves block permutation and rotation, so none of them can handle Type 1 puzzle introduced by ETCS well.
Moreover, in the security evaluation of ETCS in \cite{Kiya:ETC:TIFS19, Kiya:safe:ICASSP17, Kiya:safe:ICME17, Kiya:safeDistorsion:IWSDA17, Kiya:safe:TIF18}, Kiya {\em et al.} only considered the impact of inversion and NPT in calculation of compatibility score.
They adopted Gallagher's puzzle solver proposed in \cite{Gallagher:puzzle:CVPR12}, which cannot correctly reassemble blocks operated by inversion and NPT, even though the right matches can be recovered by using the compatibility metrics.
Figure~\ref{fig:error} illustrates two serious mistakes when applying the solver to Type 1 puzzle.
Due to lack of mechanism to deal with inverted blocks, merging two fragments of blocks gets wrong or fails.

To address the above issues, we devised an elaborate jigsaw puzzle solver tailored to Type 1 puzzle.
Essentially, a puzzle can be represented as a graph: every block is viewed as a vertex;
the quantized match degree between each pair of blocks, i.e. the compatibility score,
is regarded as a weighted edge.
Then, one can solve the puzzle by constructing a minimal spanning tree (MST) of the graph.
Our puzzle solver is built based on Kruskal's algorithm written in \cite{kruskal1956shortest}.

To facilitate the following description, we first briefly review Kruskal's algorithm.
In the beginning, each vertex is viewed as a separate tree.
Then, keep merging trees repeatedly until only one tree is left.
In each merging process, an edge with the minimum weight is selected from the set of edges.
If the two vertexes associated with the chosen edge belong to two different trees, then merge the two separate trees and record the edge;
otherwise, declare this merging operation fail and discard the edge.
Finally, all vertexes are located in one single tree, and for a graph with $n_v$ vertexes, there are $n_v-1$ recorded edges that compose the MST.
In the context of jigsaw puzzles, the tree is corresponding to a fragment built from blocks.

A puzzle can be similarly solved by repeatedly merging separate fragments.
However, the bottleneck problem is the strict geometric requirement of the jigsaw puzzle that no more than one block can occupy one single position.
Therefore, one must record the spatial positions of blocks within each fragment $F$, and check whether two fragments overlap each other during the merging process to maintain the validity of the final solution.
Besides the position of each block $u$, one needs to record its states
marking the existence of inversion and NPT, $i_u$, $t_u$, and $r_u\in [0,4]$, and $90^\circ\cdot r_u$ is its rotation angle.

The proposed puzzle solver can be described as follows:
\begin{itemize}
\item \textit{Step 1}: Assign each block $u$ to a fragment $F_u$, set its states $r_u$, $i_u$ and $t_u$ as zero, and sort the edges according to their weights.

\item \textit{Step 2}: Find the edge with the minimal weight $e^*=(u, v, s_u, s_v, i, t)$ from the edge set. Discard the edge and re-select an edge if any of the following conditions is satisfied:
\begin{itemize}
    \item The affiliated fragment of block $u$, $F_u$, and that of
    $v$, $F_v$, are the same one.

    \item Sides $s_u$ or $s_v$ are connected with other blocks.
\end{itemize}

\item \textit{Step 3}: Examine geometric validity:
    \begin{itemize}
      \item \textit{Step 3a}: Move fragment $F_v$ to place block $v$ adjacent to $u$,
        and then rotate $F_v$ so that sides $s_u$ and $s_v$ coincide.
        Denote the rotated fragment by $F'_v$.
    \item \textit{Step 3b}: If $i_u\oplus i_v\neq i$, invert $F'_v$ along the vertical direction of side $s_v$, resulting in an inverted fragment $F''_v$.
If there is any spatial collision between $F_u$ and $F''_v$, then discard $e^*$ and go back to \textit{Step 2}.
\end{itemize}

\item \textit{Step 4}: Merge $F_u$ and $F''_v$ into one fragment and update states of blocks.
For each block $a$ in $F''_v$, update $r_a$ according to the previous rotation,
and set $i_a=i_a\oplus i_u\oplus i_v\oplus i$ and $t_a=t_a\oplus t_u\oplus t_v\oplus t$.

\item \textit{Step 5}: Terminate the solver if all blocks belong to the same fragment, and otherwise go back to \textit{Step 2}.
\end{itemize}

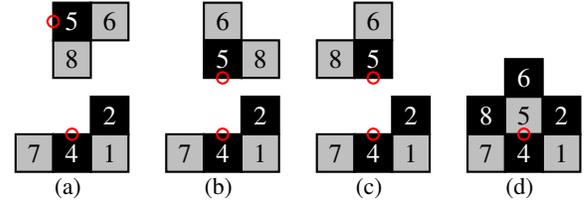
\begin{figure}[!htb]
\centering
\subfloat[]{\begin{tikzpicture}[thick=ultra thick, scale=0.5, font=\normalsize]
\foreach \index/\id in {0/7,2/1,10/8,14/6} {
  \pgfmathsetmacro\x{mod(\index,3)}
  \pgfmathsetmacro\y{\index>=6 ? floor(\index/3)-0.5 : floor(\index/3)}
  \filldraw[fill=lightgray] (\x,\y) rectangle (\x+1,\y+1);
  \node at (\x+0.5, \y+0.5) {\id};
}
\foreach \index/\id in {1/4,13/5,5/2} {
  \pgfmathsetmacro\x{mod(\index,3)}
  \pgfmathsetmacro\y{\index>=6 ? floor(\index/3)-0.5 : floor(\index/3)}
  \filldraw[fill=black] (\x,\y) rectangle (\x+1,\y+1);
  \node[white] at (\x+0.5, \y+0.5) {\id};
}
\draw[red] (1.5, 1) circle [radius=0.15];
\draw[red] (1, 4) circle [radius=0.15];
\end{tikzpicture}}
\quad
\subfloat[]{\begin{tikzpicture}[thick=ultra thick, scale=0.5, font=\normalsize]
\foreach \index/\id in {0/7,2/1,11/8,13/6} {
  \pgfmathsetmacro\x{mod(\index,3)}
  \pgfmathsetmacro\y{\index>=6 ? floor(\index/3)-0.5 : floor(\index/3)}
  \filldraw[fill=lightgray] (\x,\y) rectangle (\x+1,\y+1);
  \node at (\x+0.5, \y+0.5) {\id};
}
\foreach \index/\id in {1/4,10/5,5/2} {
  \pgfmathsetmacro\x{mod(\index,3)}
  \pgfmathsetmacro\y{\index>=6 ? floor(\index/3)-0.5 : floor(\index/3)}
  \filldraw[fill=black] (\x,\y) rectangle (\x+1,\y+1);
  \node[white] at (\x+0.5, \y+0.5) {\id};
}
\draw[red] (1.5, 1) circle [radius=0.15];
\draw[red] (1.5, 2.5) circle [radius=0.15];
\end{tikzpicture}}
\quad
\subfloat[]{\begin{tikzpicture}[thick=ultra thick, scale=0.5, font=\normalsize]
\foreach \index/\id in {0/7,2/1,9/8,13/6} {
  \pgfmathsetmacro\x{mod(\index,3)}
  \pgfmathsetmacro\y{\index>=6 ? floor(\index/3)-0.5 : floor(\index/3)}
  \filldraw[fill=lightgray] (\x,\y) rectangle (\x+1,\y+1);
  \node at (\x+0.5, \y+0.5) {\id};
}
\foreach \index/\id in {1/4,10/5,5/2} {
  \pgfmathsetmacro\x{mod(\index,3)}
  \pgfmathsetmacro\y{\index>=6 ? floor(\index/3)-0.5 : floor(\index/3)}
  \filldraw[fill=black] (\x,\y) rectangle (\x+1,\y+1);
  \node[white] at (\x+0.5, \y+0.5) {\id};
}
\draw[red] (1.5, 1) circle [radius=0.15];
\draw[red] (1.5, 2.5) circle [radius=0.15];
\end{tikzpicture}}
\quad
\subfloat[]{\begin{tikzpicture}[thick=ultra thick, scale=0.5, font=\normalsize]
\foreach \index/\id in {0/7,2/1,4/5} {
  \pgfmathsetmacro\x{mod(\index,3)}
  \pgfmathsetmacro\y{floor(\index/3)}
  \filldraw[fill=lightgray] (\x,\y) rectangle (\x+1,\y+1);
  \node at (\x+0.5, \y+0.5) {\id};
}
\foreach \index/\id in {1/4,5/2,3/8,7/6} {
  \pgfmathsetmacro\x{mod(\index,3)}
  \pgfmathsetmacro\y{floor(\index/3)}
  \filldraw[fill=black] (\x,\y) rectangle (\x+1,\y+1);
  \node[white] at (\x+0.5, \y+0.5) {\id};
}
\draw[red] (1.5, 1) circle [radius=0.15];
\end{tikzpicture}}
\caption{The process of merging two fragments that involves rotation, inversion, and NPT:
a) The two fragments to be merged;
b) Rotate the upper fragment;
c) Invert the upper fragment horizontally;
d) Apply NPT to each block in the upper fragment and then merge.}
\label{fig:merge}
\end{figure}

Note that recovering one plain-image encrypted by ETCS should be regarded as solving three individual puzzles for the YCbCr components, which are independent of each other as shown in Fig.~\ref{fig:plain256}b).
Hence, it is required that the number of blocks in $F_u$ and $F_v$ does not exceed $\lfloor\frac{n}{3}\rfloor$ in \textit{Step 2}, and the final termination condition is that all blocks belong to three different fragments.
Figure~\ref{fig:merge} illustrates the process of merging two fragments, where the circled sides of blocks are put together to form a common one, and the color of blocks indicates the state $i_t$.


\subsection{Attacking result on ETCS}
\label{ssec:attackETCS}

The following measures introduced in \cite{Gallagher:puzzle:CVPR12} are adopted in this subsection:
\begin{itemize}
\item Neighbor comparison ($N_c$):
the ratio between adjacent pairwise blocks assembled correctly and the total number.
It depends more on the performance of the pairwise compatibility metric than the assembly strategy.
For a cipher-image divided into blocks of size $3\times x\times y$, there are $3\times(2xy-x-y)$ correct pairwise blocks,
where $x=\left\lfloor \frac{W}{W_{\rm B}}\right\rfloor$ and $y=\left\lfloor\frac{H}{H_{\rm B}}\right\rfloor$.

\item Largest component ($L_c$):
the ratio between the number of blocks in the largest fragment assembled correctly and the total number.
This measure implies the size of correct local visual information contained in the recovered result.
Note that the largest fragment is selected from the three separate puzzles and the total number of blocks is
$\lfloor \frac{n}{3} \rfloor$.
\end{itemize}
To verify the performance of the proposed puzzle solver,
some experiments were performed on a number of randomly selected natural images under laboratory environment and real online network.

\subsubsection{Attacking ETCS over laboratory environment}

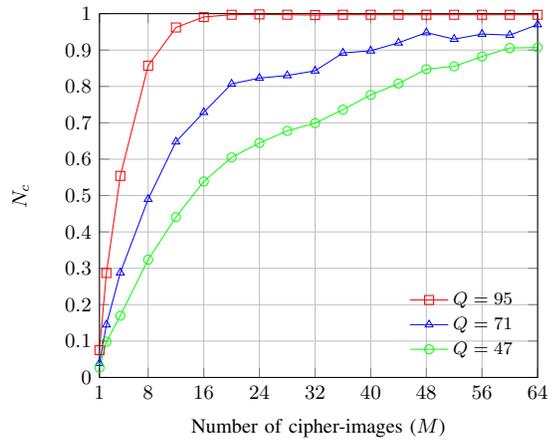
\begin{figure}[!htb]
\centering
\pgfplotstableread{
 n Q     QQ    QQQ
 1 0.075 0.039 0.02638
 2 0.287 0.145 0.09829
 4 0.554 0.288 0.16969
 8 0.857 0.490 0.32376
12 0.962 0.648 0.44069
16 0.991 0.729 0.53881
20 0.997 0.807 0.60484
24 0.998 0.823 0.64466
28 0.997 0.830 0.67776
32 0.996 0.843 0.69909
36 0.997 0.892 0.73622
40 0.997 0.898 0.77671
44 0.997 0.920 0.80796
48 0.997 0.948 0.84745
52 0.997 0.930 0.85517
56 0.997 0.944 0.88239
60 0.997 0.941 0.90575
64 0.997 0.970 0.90692
}\loadedtable
\begin{tikzpicture}[scale=0.85]
\begin{axis}[
xlabel={Number of cipher-images ($M$)},
ylabel={$N_c$},
xmin=1, xmax=64,
xtick={1, 8, 16, ..., 64},
ymin=0, ymax=1,
ytick={0, 0.1, ..., 1.01},
legend pos=south east,
xmajorgrids, ymajorgrids,
]

\addplot [red, mark=square] table[x=n, y=Q] {\loadedtable};
\addlegendentry{$Q=95$}
\addplot [blue, mark=triangle] table[x=n, y=QQ] {\loadedtable};
\addlegendentry{$Q=71$}
\addplot [green, mark=o] table[x=n, y=QQQ] {\loadedtable};
\addlegendentry{$Q=47$}

\end{axis}
\end{tikzpicture}
\caption{Decryption performance regarding the neighbor comparison.}
\label{fig:COAde-neighbor}
\end{figure}

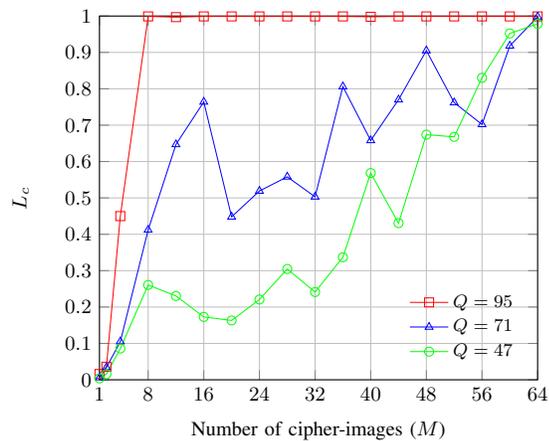
\begin{figure}[!htb]
\centering
\pgfplotstableread{
 n Q     QQ    QQQ
 1 0.016 0.006 0.00391
 2 0.036 0.034 0.01562
 4 0.450 0.105 0.08594
 8 0.999 0.412 0.26074
12 0.997 0.647 0.23047
16 0.999 0.764 0.17285
20 0.999 0.448 0.16309
24 0.999 0.519 0.22070
28 0.999 0.558 0.30469
32 0.999 0.503 0.24121
36 0.999 0.806 0.33691
40 0.998 0.658 0.56836
44 0.999 0.770 0.43066
48 0.999 0.905 0.67383
52 0.999 0.762 0.66797
56 0.999 0.702 0.83008
60 0.999 0.918 0.95215
64 0.999 0.998 0.97852
}\loadedtable
\begin{tikzpicture}[scale=0.85]
\begin{axis}[
xlabel={Number of cipher-images ($M$)},
ylabel={$L_c$},
xmin=1, xmax=64,
xtick={1, 8, 16, ..., 64},
ymin=0, ymax=1,
ytick={0, 0.1, ..., 1.01},
legend pos=south east,
xmajorgrids, ymajorgrids,
]

\addplot [red, mark=square] table[x=n, y=Q] {\loadedtable};
\addlegendentry{$Q=95$}
\addplot [blue, mark=triangle] table[x=n, y=QQ] {\loadedtable};
\addlegendentry{$Q=71$}
\addplot [green, mark=o] table[x=n, y=QQQ] {\loadedtable};
\addlegendentry{$Q=47$}

\end{axis}
\end{tikzpicture}
\caption{Decryption performance regarding the largest component.}
\label{fig:COAde-largest}
\end{figure}

\begin{figure}[!htb]
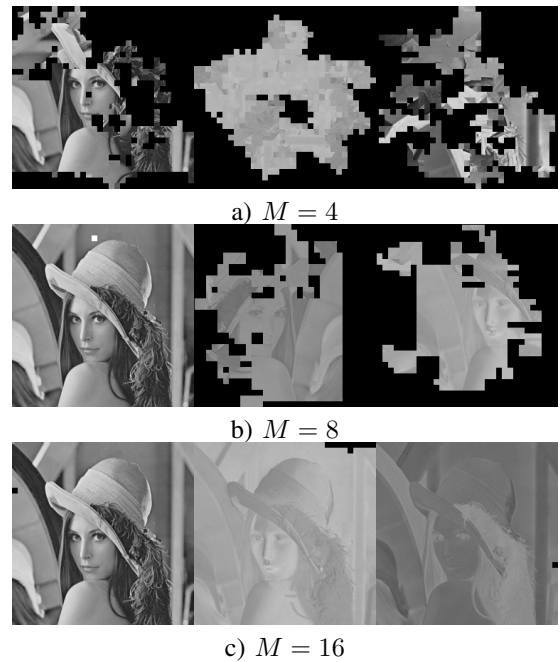

  \centering
  \begin{minipage}{\BigOneImW}
    \centering
    \includegraphics[width=\BigOneImW]{4}
    a) $M=4$
  \end{minipage}
  \begin{minipage}{\BigOneImW}
    \centering
    \includegraphics[width=\BigOneImW]{8}
    b) $M=8$
  \end{minipage}
  \begin{minipage}{\BigOneImW}
    \centering
    \includegraphics[width=\BigOneImW]{16}
    c) $M=16$
  \end{minipage}
\caption{Recovered images using $M$ cipher-images with $Q=95$.}
\label{fig:result95}
\end{figure}

\begin{figure}[!htb]
  \centering
  \begin{minipage}{\BigOneImW}
    \centering
    \includegraphics[width=\BigOneImW]{4}
    a) $M=4$
  \end{minipage}
  \begin{minipage}{\BigOneImW}
    \centering
    \includegraphics[width=\BigOneImW]{8}
    b) $M=8$
  \end{minipage}
  \begin{minipage}{\BigOneImW}
    \centering
    \includegraphics[width=\BigOneImW]{16}
    c) $M=16$
  \end{minipage}
  \begin{minipage}{\BigOneImW}
    \centering
    \includegraphics[width=\BigOneImW]{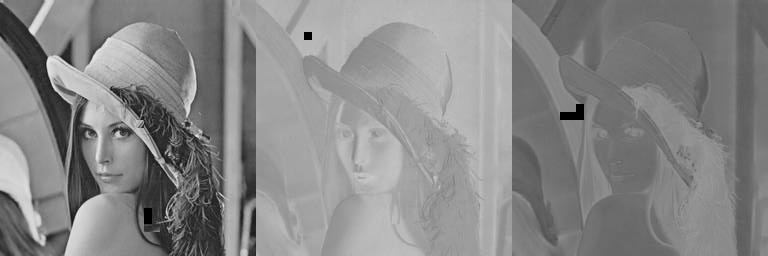}
    d) $M=72$
  \end{minipage}
  \caption{Recovered images using $M$ cipher-images with $Q=71$.}
\label{fig:result71}
\end{figure}

In local environment, the recovery performance is mainly affected by lossy compression levels and image sizes.
The compression removes some subtle information of blocks, which makes it hard to find the correct matches.
For example, when applying an extremely high-level compression to cipher-images, the pixel values within each block tend to be the same, in which an image block become a single pixel actually.
Therefore, the lower the level of lossy compression, the better the recovery performance.
As shown in Figs.~\ref{fig:COAde-neighbor} and \ref{fig:COAde-largest}, given some cipher-images encrypted by the same secret key, the assembly accuracy monotonously increases with respect to the JPEG quality factor.
For illustration, the decryption results under $Q=95$ and $Q=71$ are shown in Fig.~\ref{fig:result95} and \ref{fig:result71}, respectively.
One can see that when $M=4$ and $Q=95$, a grayscale image containing rich profile information is obtained,
which is a fragment of the luminance component.
When the number of available cipher-images reaches 16, almost all the visual information of the original plain-image can be recovered.

\graphicspath{{figures-pdf/}}

When the quality factor of cipher-images is reduced, the attacking performance becomes significantly worse, which is consistent with the result given in \cite{Kiya:safeDistorsion:IWSDA17}.
As shown in Fig.~\ref{fig:result71}c), one can only get an incomplete luminance component and two uninformative chrominance components from 16 cipher-images with $Q=75$.
Fortunately, one still can achieve an acceptable decryption result via adopting more cipher-images, as illustrated in Fig.~\ref{fig:result71}d).
In addition, it is much easier to recover the luminance component compared to the chrominance components, mainly because more subtle and distinguishable features are retained in the former after compression.

\graphicspath{{figures-pdf/result/256-95/}}

\graphicspath{{figures-pdf/result/256-71/}}

\begin{figure}[!htb]
\centering
\includegraphics[width=\OneImW]{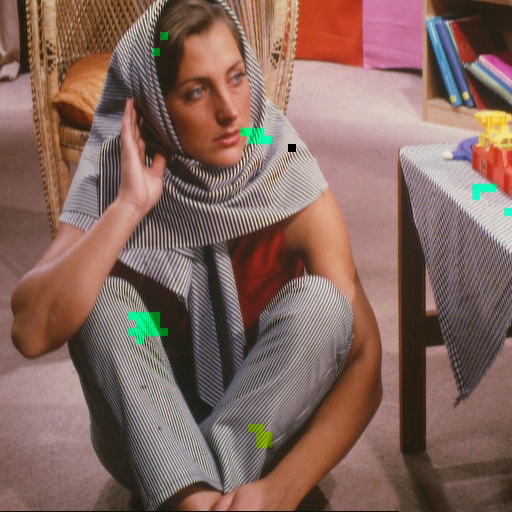}
\caption{A recovered image using the information obtained from 16 cipher-images of size $512\times 512$.}
\label{fig:result512}
\end{figure}

A large-size cipher-image contains more blocks, which causes
determination of correct matches between neighbouring blocks and
their re-assembly both much difficult.
Fortunately, the proposed puzzle solver possesses good generality, and still can work well, even though the number of blocks increases by four or 16 times.
As listed in Table~\ref{table:image-size}, there is only a marginal difference between accuracies for three image sizes when $M\geq 8$. 
Figure~\ref{fig:result512} shows a well-recovered image.
In general, the more cipher-images used in the attack, the closer $N_c$ and $L_c$ approach one.

\begin{table}[!htb]
  \centering
  \caption{Decryption performance for plain-images of various sizes.}
  \begin{tabular}{cccccc}
    \toprule
    \multirow{2}{*}{Measure} & \multirow{2}{*}{Image size} & \multicolumn{4}{c}{Number of cipher-images ($M$)} \\
    \cmidrule{3-6}
     & & 4 & 8 & 12 & 16 \\
    \midrule
    \multirow{3}{*}{$N_c$}
     & $256 \times 256$   & 0.554 & 0.857 & 0.962 & 0.991 \\
     & $512 \times 512$   & 0.496 & 0.872 & 0.979 & 0.996 \\
     & $1024 \times 1024$ & 0.726 & 0.888 & 0.935 & 0.943 \\
     \midrule
    \multirow{3}{*}{$L_c$}
     & $256 \times 256$   & 0.450 & 0.999 & 0.997 & 0.999 \\
     & $512 \times 512$   & 0.968 & 0.999 & 0.999 & 0.999 \\
     & $1024 \times 1024$ & 0.386 & 0.950 & 0.998 & 0.967 \\
    \bottomrule
  \end{tabular}
  \label{table:image-size}
\end{table}

\graphicspath{{figures-pdf/}}

\begin{figure}[!htb]
  \centering
  \includegraphics[width=\BigOneImW]{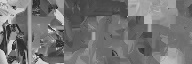}
  \caption{The recovered result for a cipher-image of size $64 \times 64$.}
  \label{fig:result64}
\end{figure}

Computational complexity is another key point of concern in an attack \cite{cqli:delay:IEEEM22, Paikin:puzzle:CVPR15}.
Actually, most time of the above attack is spent on calculating MGC scores.
Furthermore, the theoretical complexity of this part cannot be further reduced and increases exponentially with the increase of the size of cipher-images.
Fortunately, calculating MGC scores for multiple blocks can be run in parallel, so one can accelerate this part significantly via utilizing GPU computing.
To verify this point, the attacking method is implemented by Python3.8 on a PC (Intel Core i7-6850K 3.6GHz, NVIDIA GeForce GTX1080Ti).
It costs only about $30M$ seconds to solve a puzzle with 3,072 pieces, where $M$ is the number of used cipher-images.
As a comparison, the method proposed in \cite{Kiya:ETC:TIFS19} needs 166.11 minutes to solve a puzzle with 2,304 pieces ($W_{\rm B}=H_{\rm B}=8$).
It is worth noting that ETCS is not suitable for encrypting small-size images, such as thumbnails.
In such cases, more correct matches can be found effortlessly.
Furthermore, even using only one cipher-image, one can obtain some local visual information from the recovered image shown in Fig.~\ref{fig:result64}.

\subsubsection{Attacking ETCS over Online Social Networks}
\label{ssec:osn}

Sharing images privately via a social media platform is a typical application scenario of ETCS, in which the uploaded images are processed by multiple processing operations, including re-compression, resizing, and enhancement filtering, to save storage costs and improve visual experience \cite{Zhoujt:share:TOMM18}.
The resizing can be avoided through restricting the size of uploaded images to prevent severe distortion \cite{Kiya:ETC:TIFS19},
but others cannot, which inevitably changes the content of images.
Applying the operations directly on encrypted images may affect the correlation among blocks in an unexpected way.
Therefore, to verify the effectiveness of the ciphertext-only attack in a real scenario, we implemented the attack on cipher-images uploaded to Facebook (Meta) and Weibo.

\begin{figure}[!htb]
  \centering
  \begin{minipage}{\twofigwidth}
    \centering
    \includegraphics[width=\twofigwidth]{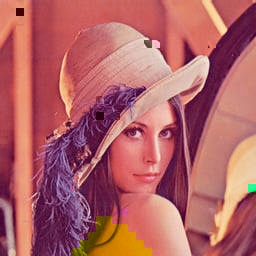}
    (a)
  \end{minipage}
    \begin{minipage}{\twofigwidth}
    \centering
    \includegraphics[width=\twofigwidth]{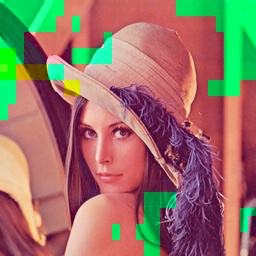}
    (b)
  \end{minipage}
  \caption{Recovered images using $M$ cipher-images downloaded from OSNs: a) Facebook ($M=64$); b) Weibo ($M=16$).}
\label{fig:OSNs}
\end{figure}

Just as in ETCS, the plain-images are first encrypted and compressed, and then uploaded to an OSN, 
in which the cipher-images went through the following operations in turn:
decompression, a series of manipulations, and re-compression.
Finally, the processed cipher-images are downloaded and then decompressed for final decryption.
The performance of attack over Facebook and Weibo are listed in Tables~\ref{table:Facebook} and \ref{table:Weibo}, respectively.
For comparison, the previous experimental results are also given.
And recovered images are shown in Fig.~\ref{fig:OSNs}.
Due to the destruction of associated information between blocks incurred by the extra operations applied to cipher-images,
the decryption becomes much more difficult.
Nonetheless, one still can obtain acceptable decryption result by using sufficient number of cipher-images.

\begin{table}[!htb]
  \centering
  \caption{Decryption performance over Facebook.}
  \begin{tabular}{cccccc}
    \toprule
    \multirow{2}{*}{Measure} & \multirow{2}{*}{Operator} & \multicolumn{4}{c}{Number of cipher-image ($M$)} \\
    \cmidrule{3-6}
     & & 16 & 32 & 48 & 64 \\
    \midrule
    \multirow{2}{*}{$N_c$}
     & Facebook         & 0.519 & 0.596 & 0.918 & 0.982 \\
     & JPEG $(Q=71)$  & 0.729 & 0.843 & 0.948 & 0.970 \\
     \midrule
    \multirow{2}{*}{$L_c$}
     & Facebook         & 0.592 & 0.276 & 0.624 & 0.996 \\
     & JPEG $(Q=71)$  & 0.764 & 0.503 & 0.905 & 0.998 \\
    \bottomrule
  \end{tabular}
  \label{table:Facebook}
\end{table}

\begin{table}[!htb]
\centering
\caption{Decryption performance over Weibo.}
  \begin{tabular}{cccccc}
    \toprule
    \multirow{2}{*}{Measure} & \multirow{2}{*}{Operator} & \multicolumn{4}{c}{Number of cipher-images ($M$)} \\
    \cmidrule{3-6}
     & & 4 & 8 & 12 & 16 \\
    \midrule
    \multirow{2}{*}{$N_c$}
     & Weibo            & 0.541 & 0.846 & 0.945 & 0.972 \\
     & JPEG $(Q=95)$  & 0.554 & 0.857 & 0.962 & 0.991 \\
     \midrule
    \multirow{2}{*}{$L_c$}
     & Weibo            & 0.460 & 0.999 & 0.999 & 0.999 \\
     & JPEG $(Q=95)$  & 0.450 & 0.999 & 0.997 & 0.999 \\
    \bottomrule
  \end{tabular}
  \label{table:Weibo}
\end{table}

\graphicspath{{figures-pdf}}

\section{Plaintext attack on ETCS}
\label{sec:PA}

The encryption scheme of ETCS is composed of two parts: inter-block permutation and intra-block manipulation.
Let $B_k(i)$ denote the possible encryption result of block $B(i)$ by the latter part, where $0\leq k\leq 15$ (considering the collision between rotation and inversion).
Then, the encryption scheme can be represented as a sequence $\mathbf{W}=\{w(i)\}_{i=0}^{n-1}$,
where $w(i)=(i', k)$ and $i'$ is the permuted location of $B(i)$.
In this section, we first present known-plaintext attack and chosen-plaintext attack based on the assumption that the attacker can generate exactly the same encryption result as the cipher-image.
Later, the attack without the assumption is also discussed.

\subsection{Known-plaintext attack on ETCS}
\label{ssec:KPA}

\begin{figure}[!htb]
\centering
\includegraphics[width=\BigOneImW]{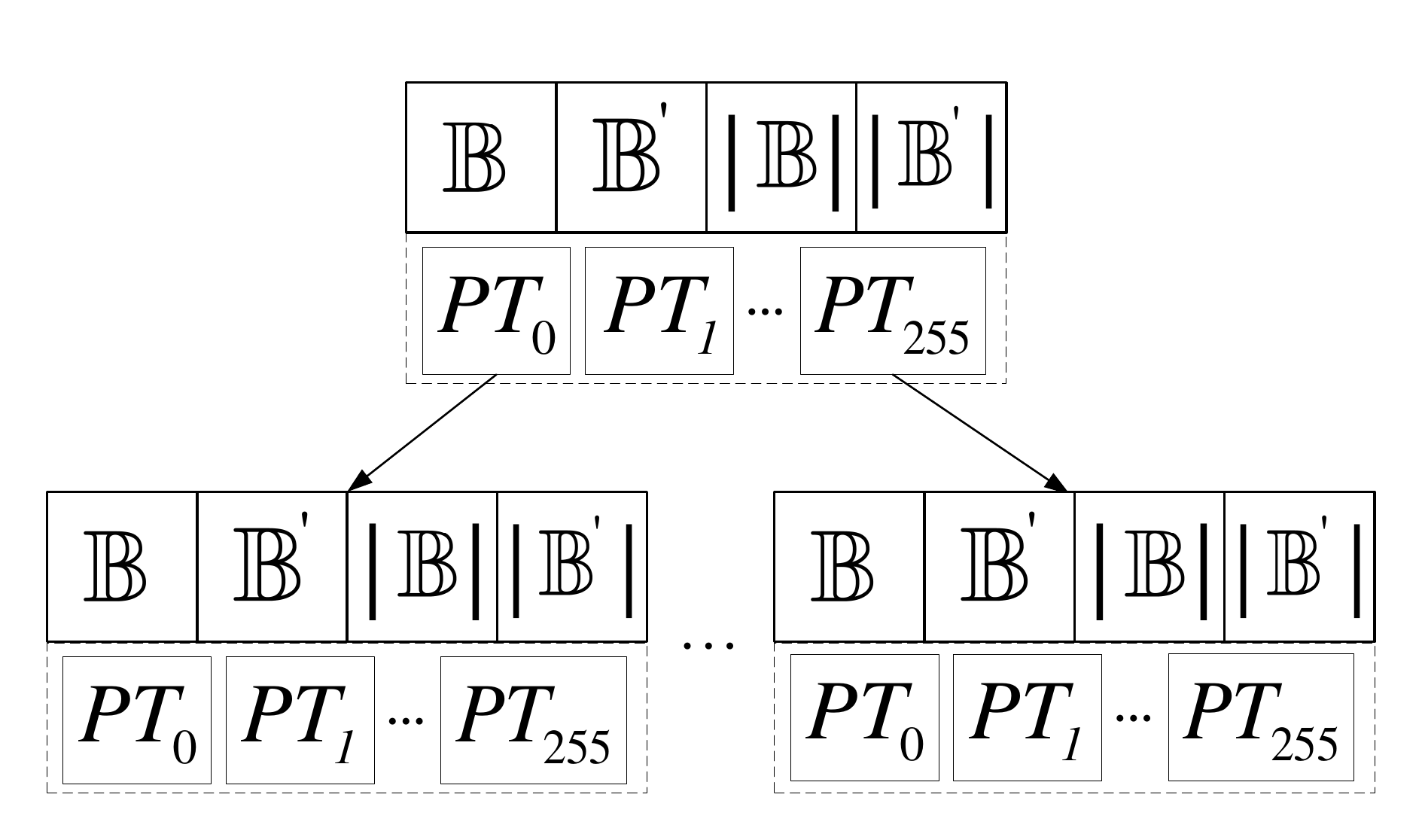}
\caption{The multi-branch tree for know-plaintext attack on ETCS.}
\label{fig:multi-branch}
\end{figure}

When plain-images and the corresponding cipher-images are available, one can generate all possible encryption results of blocks by intra-block manipulation in plain-images.
Then, the permuted locations of blocks can be determined by comparing the generated blocks with those in cipher-images pixel by pixel.
In addition, the other encryption operations applied to blocks can also be confirmed.
Similar to \cite{Cqli:Scramble:IM17}, one can retrieve $\mathbf{W}$ through building a multi-branch tree.
Each node in the tree contains five components:
a pointer array of length 256 storing addresses of its child nodes, $PT$;
a set containing the positions and types of the generated blocks, $\mathbb{B}$;  
a set containing the positions of blocks in the cipher-image, $\mathbb{B}'$;
cardinalities of the two sets. The cardinalities of $\mathbb{B}$ and $\mathbb{B}'$ are $16n$ and $n$, respectively.
The structure of the multi-branch tree is illustrated in Fig.~\ref{fig:multi-branch}.

\graphicspath{{figures-pdf/}}

Denote the $j$-th pixel of block $B_k(i)$ by $P(i, j, k)$, where $0\leq j\leq 63$.
Similarly, let $C(i, j)$ represent the $j$-th pixel of the $i$-th block $B'(i)$ in the cipher-image.
Then, the multi-branch tree can be constructed as follows.
\begin{itemize}
  \item Set $j=0$ and extend the root node with the following operations:
  \begin{itemize}
    \item $\forall\ i \in \mathbb{B}'$, add $i$ into $\mathbb{B}'_p$, where $p=C(i, j)$;

    \item $\forall\ (i, k) \in \mathbb{B}$, set $p=P(i, j, k)$, and add $(i, k)$ into $\mathbb{B}_p$ if $|\mathbb{B}'_p|\not=0$, otherwise discard $(i, k)$.

    \item To save storage, delete the two sets linked by the root node.
  \end{itemize}

  \item Search for all nodes satisfying $|\mathbb{B}|>0$ and $|\mathbb{B}'|>0$.
  Set $j=j+1$ and then expand each found node with the similar operations described above.
  Repeat this process till $j=64$.
\end{itemize}

During the construction process of the multi-branch tree,
some elements of $\mathbb{B}$ are discarded, and
the cardinality of set $\mathbb{B}$ approaches that of $\mathbb{B}'$ gradually.
After constructing the tree, one can obtain the estimated version of $\mathbf{W}$ from it.
Let $\mathbb{B}_j$ and $\mathbb{B}'_j$ denote the two sets of a leaf node.
For each leaf node, $w(i)=(i', k)$ can be uniquely determined if and only if $|\mathbb{B}_j|=1$, where $(i, k) \in \mathbb{B}_j$ and $i' \in \mathbb{B}'_j$.
Otherwise, there are $\prod_{j=1}^{c} {b \choose a}\cdot a!$ possible cases for $\mathbf{W}$, where $c$ is the number of leaf nodes and $a=|\mathbb{B}_j|$, $b=|\mathbb{B}'_j|$.

Assume that every element of the plain-images follows independent uniform distribution, and the possibility that one element in $\mathbf{W}$ can be exactly confirmed is
$p=1/(1+(16n-1)/256^m)$,
where $n$ is the number of blocks and $m$ is the number of pixels in a block.
For a color image of size $256 \times 256$, $n=3072$, $m=64$, $p$ is almost equal to one.
Theoretically, using one plain-image and the corresponding cipher-image, one can achieve a perfect decryption performance.
In fact, since the above assumption does not hold true for actual multimedia data and the effect of lossy compression is ignored,
the real decryption performance is worse than the rough estimation a little.

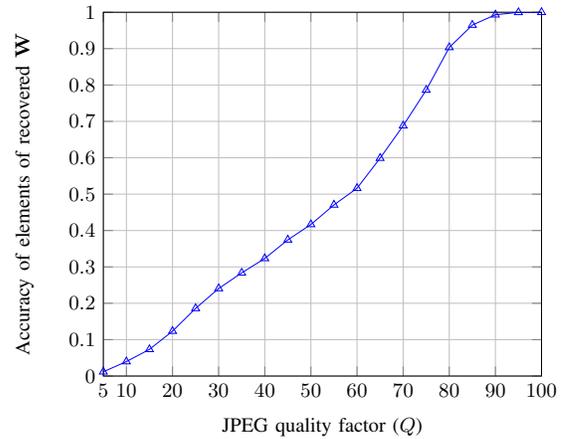
\begin{figure}[!htb]
\centering
\pgfplotstableread{
  q accuracy
  5 0.011719
 10 0.039714
 15 0.073242
 20 0.123372
 25 0.185872
 30 0.240560
 35 0.283529
 40 0.323242
 45 0.374349
 50 0.416667
 55 0.470378
 60 0.516276
 65 0.598958
 70 0.688151
 75 0.786133
 80 0.903320
 85 0.964844
 90 0.993164
 95 0.999674
100 1.000000
}\loadedtable
\begin{tikzpicture}[scale=0.85]
\begin{axis}[
xlabel={JPEG quality factor ($Q$)},
ylabel={Accuracy of elements of recovered $\mathbf{W}$},
xmin=5, xmax=100,
xtick={5, 10, 20, ..., 100},
ymin=0, ymax=1,
ytick={0, 0.1, ..., 1.01},
legend pos=south east,
xmajorgrids, ymajorgrids,
]
\addplot [blue, mark=triangle] table[x=q, y=accuracy] {\loadedtable};
\end{axis}
\end{tikzpicture}
\caption{Decryption performance of known-plaintext attack on ETCS.}
\label{fig:KPAde-rate}
\end{figure}

\begin{figure}[!htb]
  \centering
  \begin{minipage}{\ThreeImW}
    \centering
    \includegraphics[width=\ThreeImW]{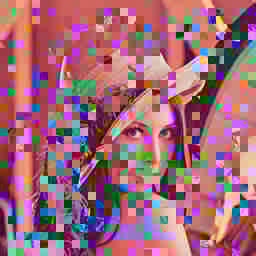}
    a) $Q=20$
  \end{minipage}
  \begin{minipage}{\ThreeImW}
    \centering
    \includegraphics[width=\ThreeImW]{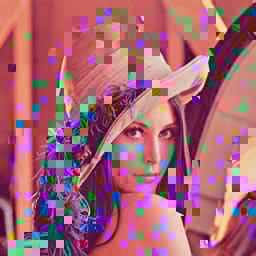}
   b) $Q=35$
  \end{minipage}
    \begin{minipage}{\ThreeImW}
    \centering
    \includegraphics[width=\ThreeImW]{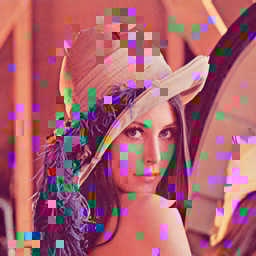}
   c) $Q=50$
  \end{minipage}
  \begin{minipage}{\ThreeImW}
    \centering
    \includegraphics[width=\ThreeImW]{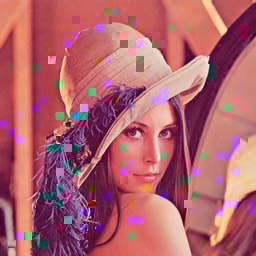}
    d) $Q=65$
  \end{minipage}
  \begin{minipage}{\ThreeImW}
    \centering
    \includegraphics[width=\ThreeImW]{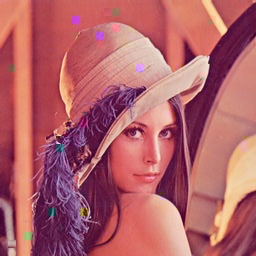}
    e) $Q=80$
  \end{minipage}
  \begin{minipage}{\ThreeImW}
    \centering
    \includegraphics[width=\ThreeImW]{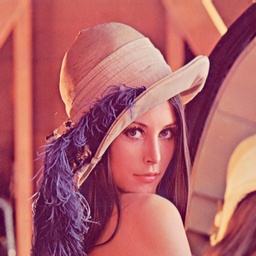}
    f) $Q=100$
  \end{minipage}
  \caption{Recovered images of various quality factors.}
  \label{fig:KPAde}
\end{figure}

Now we analyze how lossy compression affects decryption performance.
Under a small quality factor $Q$, many details of images are lost.
Then, adjacent blocks that are slightly different before compression may be the same after lossy compression. In general, the lower the quality factor, the worse the decryption performance.
Figure~\ref{fig:KPAde-rate} depicts the accuracy of the estimated version of $\mathbf{W}$, which monotonously increases regarding the JPEG quality factor.
Providing more plain-images and the corresponding cipher-images can alleviate the degradation problem.
Moreover, in practical application, the factor is generally set between 80 and 100 to obtain high visual quality,
so that a satisfying decryption accuracy ($>0.9$) can be achieved.
Figure~\ref{fig:KPAde} depicts some recovered images under various quality factors.

\graphicspath{{figures-pdf/kpa/}}

\subsection{Chosen-plaintext attack on ETCS}
\label{ssec:CPA}

As an enhanced version of known-plaintext attack, the key to chosen-plaintext attack on ETCS lies in recovering
$\mathbf{W}$ much accurately by choosing some special plain-images and observing the corresponding cipher-images.
To uniquely determine each element of $\mathbf{W}$, the cardinality of set $\mathbb{B}$ of each leaf node should be one, so all the encryption results of blocks in the chosen plain-image are required to be different.
To avoid the possible effect of conversion error between color space and irreversible loss of JPEG compression,
one should construct the plain-image in RGB space, apply color conversion and compression to it, and then verify whether it satisfies the aforementioned condition.
This ensures that all lossy operations are run in only one direction.
The procedure of constructing the chosen plain-image from a randomly generated image can be described as follows.
\begin{itemize}
\item \textit{Step 1}: Randomly generate $\lfloor\frac{n}{3}\rfloor$ blocks
of size $W_{\rm B} \times H_{\rm B} \times 3$ in RGB color space.

\item \textit{Step 2}: Transform every block into YCbCr color space, and apply compression and decompression to them. Then split each block into three blocks of size $W_{\rm B} \times H_{\rm B}$.

\item \textit{Step 3}: Go through each divided block and record its 16 encryption results.
If any result of a block is the same as that previously recorded one, then remove the block
generating this block.

\item \textit{Step 4}: If the number of removed blocks in \textit{Step 3} is greater than zero, then generate the same number of blocks in RGB space and go back to \textit{Step 2}.
Otherwise, combine the remaining blocks in RGB space and form a chosen plain-image of size $W\times H$.
\end{itemize}

\begin{figure}[!htb]
 \centering
  \begin{minipage}{\ThreeImW}
    \centering
    \includegraphics[width=\ThreeImW]{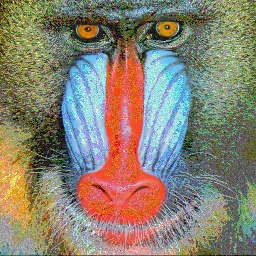}
    a)
  \end{minipage}
  \begin{minipage}{\ThreeImW}
    \centering
    \includegraphics[width=\ThreeImW]{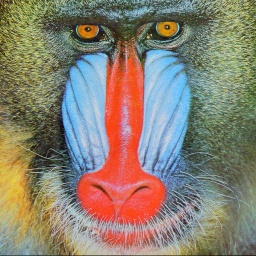}
    b)
  \end{minipage}
  \begin{minipage}{\ThreeImW}
    \centering
    \includegraphics[width=\ThreeImW]{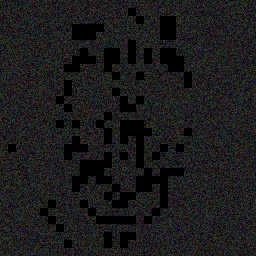}
    c)
  \end{minipage}
  \caption{An example of chosen plain-image: a) the chosen plain-image; b) the original image; c) the result of bitwise XOR between a) and b).}
  \label{fig:CPA}
\end{figure}

Besides constructing the plain-images from a noisy image, one can also
produce them by refining a natural image for the attack.
Similar to the above procedure, one should modify the pixels in blocks that do not satisfy the expected condition.
To avoid serious visual change, one can just alter the least significant bits of the pixel values.
Figure~\ref{fig:CPA} shows the modified image, its original image, and the result of bitwise XOR between them.
It can be seen that there is only a slight visual difference between them.
Note that when the quality factor decreases, the difference becomes more apparent.

\graphicspath{{figures-pdf}}

\subsection{Plaintext attack on ETCS over Online Social Networks}
\label{ssec:pc:sim}

In some cases, the attacker cannot generate the same encryption result as the cipher-image
due to various reasons, such as the quantization tables employed in compression are different;
the attacker may not be able to know the details of the manipulations operated by an OSN on the cipher-image.
Under such circumstances, the elements of $\mathbf{W}$ can be determined through comparing the similarity between blocks.
For simplicity, the similarity between blocks $u$ and $v$ is measured by the sum of squared distance:
\begin{equation*}
D(u, v)=\sum_{i=0}^{W_{\rm B}-1}\sum_{j=0}^{H_{\rm B}-1}(u(i, j)-v(i, j))^2.
\end{equation*}

To perform known-plaintext attacks, one needs to generate all the possible encryption results for each block of the plain-image.
The generated blocks and those in the cipher-image are stored in two sets, $\mathbb{P}$ and $\mathbb{C}$, respectively.
Although the previously generated block and its corresponding one in $\mathbb{C}$ are not exactly the same, it can be assumed that their similarity is maximum.
According to the calculated distances, the elements of $\mathbf{W}$ can be greedily recovered.
Each time, two blocks $p$ and $c$ with the smallest distance are selected from $\mathbb{P}$ and $\mathbb{C}$, respectively.
Remove the blocks coming from the same block as $p$ from the set $\mathbb{P}$.
Continue this process till the two sets are empty.
Finally, $\mathbf{W}$ can be recovered based on the relationship between the selected blocks.
When the image shown in Fig.~\ref{fig:plain256}a) is used as a plain-image, the ratios of correctly recovered elements of $\mathbf{W}$ are 0.634 and 1.0 in attacks implemented on platforms Facebook and Weibo, respectively.

\section{Security of a conventional ETCS}
\label{sec:conv}

There are several conventional ETCSs proposed in \cite{Kiya:ETCJPEG:PCS15, Kiya:ETCMJPEG:TFE15, Kiya:ETCJPEGXR:BMSB16, Kiya:ETCLossless:TIS17, Kiya:ETCRGB:SCIA17, Kiya:ETCRGB:TIS2018},
which are called as ETC in this paper for brevity.
Some of them designed for JPEG-LS \cite{Kiya:ETCRGB:SCIA17, Kiya:ETCRGB:TIS2018} can be considered equivalent versions of ETCS that produces color cipher-images, so the previous analysis applies to them as well.
The others are designed for JPEG, JPEG XR, and Motion JPEG, some of which can enable lossless compression.
Their essential encryption processes are equivalent, so it can be inferred that their security performances are almost the same.
Since it is found that the distortion incurred by lossy compression has a detrimental effect on decryption,
we focus on the lossy JPEG compression in the section and investigate the performance of the proposed attacking method in this worse situation.

\subsection{Defects of ETC}
\label{ssec:conv:defect}

In ETC, the entire encryption is performed directly in RGB space and the cipher-image is still a color image.
After permutation, block rotation and inversion, and NPT,
the color components within blocks are shuffled to enhance the system security.
Specifically, three color components \textit{RGB} of each block are randomly permuted as one of six possible results: \textit{RGB, RBG, BGR, BRG, GBR, GRB}.

\begin{figure}[!htb]
\centering
\includegraphics[width=\OneImW]{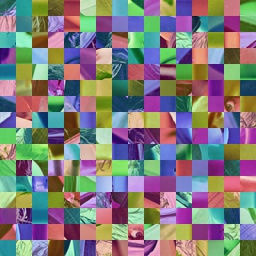}
\caption{A cipher-image of the ETC.}
\label{fig:cipher-conv}
\end{figure}

The defects of ETC with lossy JPEG compression are not only manifested in compression stage but also in decompression stage.
In compression, there is an operation of chrominance sub-sampling, which strictly limits the block size used in encryption to $16\times 16$.
More chrominance information is discarded in the compression of color images to further improve compression ratio, considering that human beings are much more sensitive to changes in luminance than chrominance.
In 4:2:0 sub-sampling scheme, the chrominance size of a color image of size $W\times H$ is reduced to $\lfloor\frac{W}{2}\rfloor\times\lfloor\frac{H}{2}\rfloor$.
Thus, the compression is performed inside each block of size $8\times 8$, which is derived from an encrypted block of size $16\times 16$.
If one chooses a block size less than $16\times 16$ in encryption, then the pixels of each block in compression come from some encrypted ones of low correlation, causing that the compression performance drops sharply.
However, the larger size obviously weakens the security and the encryption might not be able to conceal visual information well.
As illustrated in Fig.~\ref{fig:cipher-conv}, some regional visual information is leaked in the cipher-image.
In addition, other sub-sampling schemes except 4:4:4, such as 4:2:2, can also cause the above defect.

\begin{figure}[!htb]
  \centering
  \includegraphics[width=\OneImW]{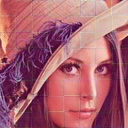}
  \caption{A decrypted image containing distortion incurred by encryption.}
\label{fig:distortion}
\end{figure}

In the decompression stage, there exists incorrect interpolation because of encryption, which causes severe distortion along the boundary of blocks in decrypted images, as shown in Fig.~\ref{fig:distortion}.
When 4:2:0 sub-sampling is applied to cipher-images in compression, the size of chrominance components should be doubled in both the horizontal and vertical directions during decompression, which is achieved via interpolation.
However, the images are not decrypted at this moment.
As a result, during the interpolation, the relationship between irrelevant blocks is employed, leading to some wrong interpolation.
To avoid this defect, the decryption should be placed inside the decompression process,
but this violates the design philosophy of ETC.
Furthermore, if the cipher-images are uploaded to an OSN,
the decompression and recompression are compulsively applied to cipher-images.
Therefore, such kind of distortion cannot be prevented in ETC and considerably reduces its usability.

\subsection{Ciphertext-only attack on ETC}

The ciphertext-only attack on ETC is similar to that on the grayscale-like ETCS.
And the MGC metric and puzzle solver described in Sec.~\ref{sec:COA} can also be employed in the attack.
The additional shuffling operation of color components can be processed like NPT.
So the details of the attack method are omitted here.

\begin{figure}[!htb]
\centering
\pgfplotstableread{
 n    color     gray   ignore
 1 0.069792 0.121724 0.121875
 2 0.108333 0.332325 0.276042
 3 0.212500 0.500168 0.445833
 4 0.278125 0.537718 0.594792
 5 0.388542 0.697581 0.741667
 6 0.443750 0.747228 0.722917
 7 0.487500 0.781082 0.808333
 8 0.510417 0.820060 0.847917
 9 0.534375 0.865171 0.867708
10 0.554167 0.881972 0.887500
11 0.589583 0.898185 0.916667
12 0.609375 0.909526 0.930208
13 0.629167 0.913726 0.926042
14 0.646875 0.925823 0.951042
15 0.654167 0.934476 0.958333
16 0.661458 0.942876 0.963542
17 0.658333 0.945481 0.964583
18 0.679167 0.948421 0.967708
19 0.672917 0.953713 0.970833
20 0.672917 0.955393 0.967708
21 0.676042 0.956485 0.967708
22 0.680208 0.962030 0.968750
23 0.678125 0.963962 0.968750
24 0.690625 0.968246 0.970833
25 0.697917 0.969590 0.971875
26 0.703125 0.969842 0.972917
27 0.713542 0.973034 0.971875
28 0.715625 0.971942 0.976042
29 0.720833 0.972614 0.976042
30 0.718750 0.974210 0.976042
31 0.728125 0.975050 0.979167
32 0.729167 0.976815 0.976042
}\loadedtable
\begin{tikzpicture}[scale=0.85]
\begin{axis}[
xlabel={Number of cipher-images ($M$)},
ylabel={Accuracy},
xmin=1, xmax=32,
xtick={1, 4, 8, ..., 64},
ymin=0, ymax=1,
ytick={0, 0.1, ..., 1.01},
legend pos=south east,
xmajorgrids, ymajorgrids,
]

\addplot [red, mark=x] table[x=n, y=color] {\loadedtable};
\addlegendentry{ETC}
\addplot [green, mark=triangle] table[x=n, y=gray] {\loadedtable};
\addlegendentry{Grayscale-like ETCS}
\addplot [blue, mark=10-pointed star] table[x=n, y=ignore] {\loadedtable};
\addlegendentry{ETC (ignoring color)}

\end{axis}
\end{tikzpicture}
\caption{Accuracy of MGC metric for ciphertext-only attack on ETC.}
\label{fig:mgc-conv}
\end{figure}
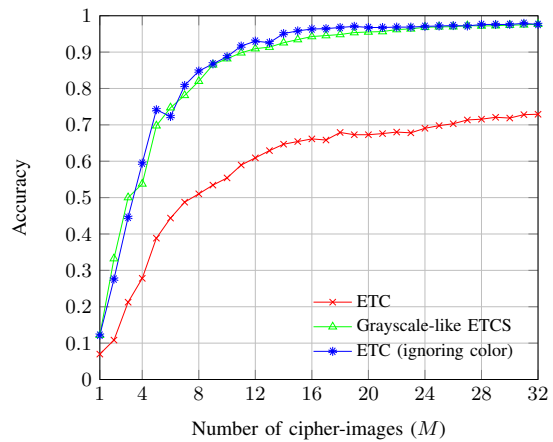

Similar to Sec.~\ref{sec:COA:pc}, we calculated accuracies of MGC metric to evaluate its effectiveness for ETC.
The calculation of MGC mainly relies on the pixels at the edge of blocks, in which the distortion occurs.
Consequently, the capability of MGC is affected as shown in Fig.~\ref{fig:mgc-conv}.
Despite the large block size and more color information of images in ETC, the corresponding accuracy is severely degraded.
To investigate the effect of the shuffle operation, we also calculated the accuracy when ignoring whether the color components between blocks are configured correctly, which is referenced as ``ignoring color'' in Fig.~\ref{fig:mgc-conv}.
The accuracy increases a lot and is comparable to that of the grayscale-like ETCS.
Therefore, one can deduce that it is difficult to find the correct configuration of color components, but it has less impact on recovering other relationships between blocks, which is also reflected in the decryption result.

The attacking performance of ETC is shown in Table~\ref{table:image-conv}.
Note that the calculation of $N_c$ and $L_c$ also ignores whether the color components are configured correctly.
For a color image of size $256\times 256$, there are $3072=3\cdot\frac{256}{8}\cdot\frac{256}{8}$ and $256=\frac{256}{16}\cdot\frac{256}{16}$ blocks for grayscale-like ETCS and ETC, respectively.
Although there are the larger block size and fewer number of blocks for ETC, the decryption performance regarding the two ETCSs is similar due to the distortion.

\begin{table}[!htb]
  \centering
  \caption{Decryption performance comparison with $Q=95$.}
  \begin{tabular}{cccccc}
    \toprule
    \multirow{2}{*}{Measure} & \multirow{2}{*}{Encryption} & \multicolumn{4}{c}{Number of cipher-images ($M$)} \\
    \cmidrule{3-6}
     & & 4 & 8 & 12 & 16 \\
    \midrule
    \multirow{2}{*}{$N_c$}
     & ETC    & 0.298 & 0.825 & 0.975 & 0.996 \\
     & Grayscale-like ETCS & 0.554 & 0.857 & 0.962 & 0.991 \\
     \midrule
    \multirow{2}{*}{$L_c$}
     & ETC    & 0.230 & 0.863 & 0.992 & 0.996 \\
     & Grayscale-like ETCS & 0.450 & 0.999 & 0.997 & 0.999 \\
    \bottomrule
  \end{tabular}
  \label{table:image-conv}
\end{table}

\begin{figure}[!htb]
  \centering
  \begin{minipage}{\twofigwidth}
    \centering
    \includegraphics[width=\twofigwidth]{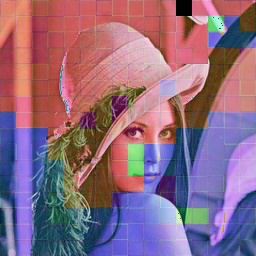}
    a)
  \end{minipage}
  \begin{minipage}{\twofigwidth}
    \centering
    \includegraphics[width=\twofigwidth]{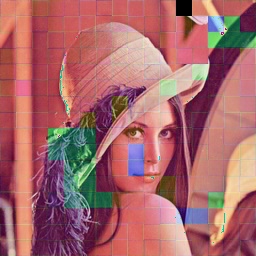}
    b)
  \end{minipage}
\caption{Recovered images using 12 cipher-images with $Q=95$:
a) the original recovered image; b) the enhanced recovered image.}
\label{fig:result-conv}
\end{figure}

As mentioned before, it is much more difficult to restore the correct configuration of color components.
As shown in Fig.~\ref{fig:result-conv}a), although most blocks are placed correctly, the recovery of color components is seriously wrong.
In the assembly process, the wrong relationship of color components is used when large fragments are merged, resulting in color differences between large areas.
To mitigate this problem, one can use a sample strategy, majority voting, which is effective and less time-consuming.
When two fragments are merged, one can find all pairs of blocks to be joined together, and then select the configuration of color components that can satisfy the most demand of the pairs.
As shown in Fig.~\ref{fig:result-conv}b), the result of modified assembly strategy becomes better.

\graphicspath{{figures-pdf/conventional/}}

\subsection{Plaintext attack on ETC}
\label{ssec:conv:PA}

\begin{figure}[!htb]
  \centering
  \begin{minipage}{\twofigwidth}
    \centering
    \includegraphics[width=\twofigwidth]{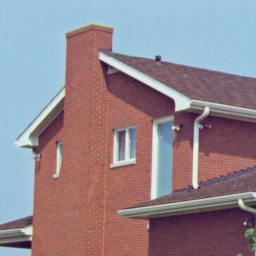}
    a)
  \end{minipage}
  \begin{minipage}{\twofigwidth}
    \centering
    \includegraphics[width=\twofigwidth]{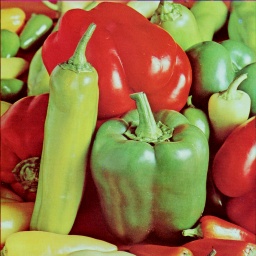}
    b)
  \end{minipage}
\caption{Two plain-images of size $256\times 256$.}
\label{fig:plain-conv}
\end{figure}

Due to the incorrect interpolation mentioned in Sec.~\ref{ssec:conv:defect}, the attacker cannot generate exactly the
same encryption result as the cipher-image.
So the plaintext attack can only be achieved via comparing the similarity between blocks, which is similar to Sec.~\ref{ssec:pc:sim}.
The sum of squared distance between blocks $u$ and $v$ in ETC is calculated via
\begin{equation*}
D(u, v)=\sum_{i=0}^{W_{\rm B}-1}\sum_{j=0}^{H_{\rm B}-1}\sum_{k=0}^{2}(u(i, j, k)-v(i, j, k))^2.
\end{equation*}
Apparently, the attacking performance is directly related to the plain-image. For example, when the two plain-images shown in Fig.~\ref{fig:plain-conv}a) and b) are employed in attack, the ratios of correctly recovered elements of $\mathbf{W}$ are 0.684 and 0.953, respectively.
Since some blocks in the plain-image shown in Fig.~\ref{fig:plain-conv}a) are very similar to each other, many elements of $\mathbf{W}$ are recovered by mistake.
Therefore, when proceeding chosen-plaintext attacks on ETC, the attacker should ensure that the similarity between blocks in the chosen plain-image is minimized.

\graphicspath{{figures-pdf/conventional/}}

In this section, we extended the ciphertext-only attack to some conventional ETCSs by making minor modifications to the puzzle solver.
It should be noted that for encryption schemes like ETCS only involving inter-block permutation and intra-block operations,
if the correlation among blocks cannot be eliminated by encryption, the schemes are considered to be vulnerable to this kind of attacks.
In this case, the correlation can be accurately estimated by machine learning methods \cite{Bridger:puzzle:CVPR2020, Paumard:puzzle:TIP2020}, which can help to determine the correct relationship between blocks.
Then, the blocks can be properly assembled by some sort of jigsaw puzzle solver.

\section{Conclusion}

This paper re-evaluated the security performance of some encryption-then-compression systems.
Using the preserved correlation between blocks in the cipher-images, we proposed a puzzle solver
to reassemble the original plain-image.
Both theoretical analysis and experimental results are provided to report that
the studied systems are still vulnerable against ciphertext-only attack, known-plaintext attack and chosen-plaintext attack. In addition, we also analyzed the insecurity of the conventional ETC against the three classic attack models. The insecurity of such systems indicates that much work is needed to deal with the balancing point among image compression, security, and usability for a given scenario.

\bibliographystyle{IEEEtran.bst}
\bibliography{ETC}
\end{document}